\documentclass[12pt]{article}

\global\arraycolsep=1pt

\usepackage{color,amsbsy,amssymb,latexsym,amsfonts, amsmath,cancel}
\usepackage{braket}
\usepackage{mathrsfs}
\usepackage{graphicx}
\usepackage{cite}

\numberwithin{equation}{section}
\newcommand{\bel}[1]{\begin{equation}\label{#1}}                     
\newcommand{\bal}[1]{\begin{eqnarray}\label{#1}}   
\newcommand{\be}{\begin{equation}}               
\newcommand{\ba}{\begin{eqnarray}}           
\newcommand{\ee}{\end{equation}}
\newcommand{\ea}{\end{eqnarray}}

\renewcommand{\thefootnote}{\fnsymbol{footnote}}

\newcommand{\bea}{\begin{equation}}
\newcommand{\eea}{\end{equation}}
\newcommand\fD{\mathfrak D}

\begin{document}

%
%
\begin{titlepage}
\begin{flushright}
\normalsize
~~~~
OCU-PHYS 380\\
\end{flushright}

\vspace{15pt}

\begin{center}
{\LARGE 
$D$-term Triggered Dynamical Supersymmetry Breaking }
\end{center}

\vspace{23pt}

\begin{center}
{ H. Itoyama$^{a, b} $\footnote{e-mail: itoyama@sci.osaka-cu.ac.jp}
  and 
 Nobuhito Maru$^c$\footnote{e-mail: maru@phys-h.keio.ac.jp} 
}\\
%
\vspace{18pt}
%

$^a$ \it Department of Mathematics and Physics, Graduate School of Science\\
Osaka City University  and\\
\vspace{5pt}

$^b$ \it Osaka City University Advanced Mathematical Institute (OCAMI) \\

\vspace{5pt}

3-3-138, Sugimoto, Sumiyoshi-ku, Osaka, 558-8585, Japan \\

$^c$ \it Department of Physics, and
Research and Education Center for Natural Sciences, \\
Keio University, Hiyoshi, Yokohama, 223-8521 JAPAN

\end{center}
%
\vspace{20pt}
\begin{center}
Abstract\\
\end{center}
%
We present the mechanism of the dynamical supersymmetry breaking at the metastable vacuum 
recently uncovered in the ${\cal N}=1$ $U(N)$ supersymmetric gauge theory 
that contains adjoint superfields and that is specified by K\"{a}hler and non-canonical gauge kinetic functions 
and a superpotential whose tree vacua preserve ${\cal N}=1$ supersymmetry.
The overall $U(1)$ serves as the hidden sector and no messenger superfield is required. 
The dynamical supersymmetry breaking is triggered by the non-vanishing $D$ term coupled to the observable sector, 
and is realized by the self-consistent Hartree-Fock approximation of the NJL type 
while it eventually brings us the non-vanishing $F$ term as well. 
It is shown that theoretical analysis is resolved as a variational problem of the effective potential for three kinds of background fields, 
namely, the complex scalar, and the two order parameters $D$ and $F$ of supersymmetry,  
the last one being treated perturbatively. We determine the stationary point and
numerically check the consistency of such treatment as well as the local stability of the scalar potential. 
The coupling to the ${\cal N} = 1$ supergravity is given and the gravitino mass formula is derived.

\vfill

\setcounter{footnote}{0}
\renewcommand{\thefootnote}{\arabic{footnote}}

\end{titlepage}

\renewcommand{\thefootnote}{\arabic{footnote}}
\setcounter{footnote}{0}

\section{Introduction}
Spontaneous breaking of rigid supersymmetry occurs much less frequent compared with that of internal symmetry
 in quantum field theory and has attracted much interest \cite{Witten1, Witten2} of theorists for over the three decades.
Mass hierarchy in elementary particle physics indicates that it is most desirable to break ${\cal N}=1$ supersymmetry dynamically. 
In fact, under the non-renormalization theorem \cite{GRS}, no holomorphic operator is generated in perturbation theory 
and instanton generated nonperturbative superpotentials have been the major source of dynamical supersymmetry breaking (DSB).
  
In this paper, we focus our attention on general rigid ${\cal N}=1$ theory  in four spacetime dimensions 
consisting of vector superfields and chiral superfields in the adjoint representation 
which permits a non-canonical gauge kinetic function $\tau_{ab}$ (that may follow from the second derivative of the prepotential) 
and hence the D term-gaugino-matter fermion (or D term-Dirac gauginos) nonrenormalizable coupling. 
It has recently been shown in refs. \cite{imaru1, imaru2} that, in this general situation, 
supersymmetry is dynamically broken in the metastable vacuum. 
The mechanism that triggers the DSB is the condensate of the Dirac bilinear above, 
forcing one of the order parameters $D$ of supersymmetry to be non-vanishing. 
This is very much reminiscent of the Nambu-Jona Lasinio (NJL) theory \cite{NJL1, NJL2} of broken chiral symmetry 
and hence the BCS superconductivity \cite{BCS, NambuPRBCS2}, 
being formulated in terms of the effective action of the auxiliary field whose stationary value is the order parameter.
The method of approximation employed is the self-consistent Hartree-Fock approximation 
where the tree and the one-loop contributions are regarded as comparable.
Once this mechanism operates, non-vanishing F term is shown to be induced and contributes, for instance, to the mass of the fermions. 
The mechanism requires massive adjoint scalars, in particular, the scalar gluons and, 
together with the feature that the D term triggers the breaking, is quite distinct from the previous proposals \cite{ADS1, ADS2, ADS3, Veneetal1, Veneetal2} of DSB 
both from theoretical and experimental perspectives. 
The overall $U(1)$ where the non-vanishing $D$ and the Nambu-Goldstone fermion (NGF) reside serves as the hidden sector 
and no messenger field is necessary \cite{imaru1} as non-vanishing third prepotential derivatives connect the $U(1)$ sector 
with the observable $SU(N)$ sector \cite{FIS1, FIS346}. 

While our treatment of the theory bears much resemblance with that of the NJL theory, 
there is one important complexity which has no counterpart in the NJL  and which we did not emphasize in ref. \cite{imaru1}. 
In NJL, aside from the pseudoscalar auxiliary massless singlet field commonly denoted by $\pi$,  
there is only one singlet auxiliary scalar field denoted by $\sigma$ in the effective action, which is the order parameter
of chiral symmetry. (See appendix A.)
In other words,  the stationary condition of energy with respect to the scalar is at the same time 
the stationarity with respect to the order parameter (the gap equation).
This is not the case here. 
After treating the $U(N)$ singlet real auxiliary field $F$ as perturbation, 
we have two kinds of background fields in the effective potential: 
these are the singlet complex scalar $\varphi$ and the singlet auxiliary field $D$. 
The stationarity of energy with respect to the scalar and that with respect to the order parameter are one 
and the other and both must be imposed simultaneously. 
In this paper, we will mainly deal with such multi-variable variational problem in depth 
and present the solution which is the local minimum of the scalar potential. 
We will also include a few other materials which have phenomenological implications. 
We work in the unbroken phase of $U(N)$ and invoke $U(N)$ invariance of the expectation values 
to suppress indices often. 
  
In the next section, we start out from exhibiting the component action from that of the superspace, 
state the set of assumptions we have made in \cite{imaru1, imaru2} and in this paper and 
give the Noether current associated with ${\cal N}=1$ rigid supersymmetry. 
We review the original reasoning that has led us to the D-term triggered dynamical supersymmetry breaking. 
We set up the background field formalism to be used in the subsequent sections, 
separating the three kinds of background from the fluctuations. 
The action can be coupled to ${\cal N}=1$ supergravity and we derive the gravitino mass formula 
via the super-Higgs mechanism associated with the non-vanishing D-term. 
The action contains a sequence of special cases 
in which the gauge coupling function and the superpotential are related in a specific form, 
including the one where the rigid ${\cal N}=2$ supersymmetry is partially broken to ${\cal N}=1$ at the tree level \cite{APT, FIS1}. 
In section three, we elaborate upon our treatment of the effective potential with the three kinds of background fields 
as well as the point of the Hartree-Fock approximation in refs. \cite{imaru1, imaru2}. 
Section four is the main thrust of this paper. We present our variational analyses of the effective potential in full detail. 
Treating one of the order parameters $F$ as an induced perturbation, 
we demonstrate that the stationary values $(D_*, \varphi_*, \bar{\varphi_*})$ are determined by the intersection of the two real curves, 
namely, the simultaneous solution to the gap equation and the equation of $\varphi$ stationarity (the energy condition). 
Numerical analysis is provided that demonstrates the existence of such solution as well as the self-consistency of our analysis. 
The second variation of the scalar potential is computed and the local stability of the vacuum is shown from the numerical data. 
We finish our paper with summary and brief comments on the issue of regularization and subtraction schemes.
   
   In two of the appendix on rudimentary materials to be referred to in the text, 
   we take a brief look at the NJL effective action and recall a formula of the second variation of a multivariable function.
   Phenomenological applications of our finding and the estimate of the longevity of
   our metastable vacuum have been given in \cite{imaru1, imaru2}, which we do not repeat in this paper.  
  
\section{The action, assumptions and some properties}
The action we work with in this paper is the general ${\cal N}=1$ supersymmetric action 
consisting of chiral superfield $\Phi^a$ in the adjoint representation 
and the vector superfield $V^a$ with three input functions, 
the K\"{a}hler potential $K(\Phi^a, \bar{\Phi}^a)$ with its gauging, the gauge kinetic superfield $\tau_{ab}(\Phi^a)$
 that follow from the second derivatives of a generic holomorphic function ${\cal F}(\Phi^a)$,
and the superpotential $W(\Phi^a)$.  
     \ba
    {\cal L}
     &=&     
           \int d^4 \theta K(\Phi^a, \bar{\Phi}^a) + (gauging)   
        + \int d^2 \theta
           {\rm Im} \frac{1}{2} 
           \tau_{ab}(\Phi^a)
           {\cal W}^{\alpha a} {\cal W}^b_{\alpha}
            + \left(\int d^2 \theta W(\Phi^a)
         + c.c. \right).       
           \label{KtauW}
    \ea
 The gauge group is taken to be $U(N)$ and, for simplicity, 
 we assume that the theory is in the unbroken phase of the entire gauge group, 
 which can be accomplished by tuning the superpotential. 
 We also assume that third derivatives of ${\cal F}(\Phi^a)$ at the scalar vev's are non-vanishing. 
\subsection{action and component expansion}

The component Lagrangian of eq. (\ref{KtauW}) reads 
\ba
{\cal L}_{U(N)} = {\cal L}_{{\rm K\ddot{a}hler}} + {\cal L}_{{\rm gauge}} + {\cal L}_{{\rm sup}},
\ea
where
\ba
 {\cal L}_{{\rm K\ddot{a}hler}}
    &=&    g_{ab} {\cal D}_\mu \phi^a {\cal D}^\mu \bar{\phi}^b
         - \frac{i}{2} g_{ab} \psi^a \sigma^\mu {\cal D}_\mu' \bar{\psi}^b
         + \frac{i}{2} g_{ab} {\cal D}_\mu' \psi^a \sigma^\mu \bar{\psi}^b + g_{ab} F^a \bar{F}^b
           \nonumber \\
    & &  - \frac{1}{2} g_{ab,\bar{c}} F^a \bar{\psi}^b \bar{\psi}^c
         - \frac{1}{2} g_{bc,a} \bar{F}^c \psi^a \psi^b
         + \frac{1}{\sqrt{2}} g_{ab} (\lambda^c \psi^a k_c^*{}^{b} 
         + \bar{\lambda}^c \bar{\psi}^b k_c{}^{a})
         + \frac{1}{2} D^a \fD_a,
           \label{L:K} \\
  {\cal L}_{{\rm gauge}}
    &=&  - \frac{1}{2} {\cal F}_{ab} \lambda^a \sigma^\mu {\cal D}_\mu \bar{\lambda}^b
         - \frac{1}{2} \bar{{\cal F}}_{ab} {\cal D}_\mu \lambda^a \sigma^\mu \bar{\lambda}^b
         - \frac{1}{4} (\Im {\cal F})_{ab} F_{\mu \nu}^a F^{b \mu \nu}
         - \frac{1}{8} (\Re {\cal F})_{ab} \epsilon^{\mu \nu \rho \sigma} 
           F_{\mu \nu}^a F_{\rho \sigma}^b
           \nonumber \\
    & &  - \frac{\sqrt{2} i}{8} ({\cal F}_{abc} \psi^c \sigma^\nu \bar{\sigma}^\mu \lambda^a
         - \bar{{\cal F}}_{abc} \bar{\lambda}^a \bar{\sigma}^\mu \sigma^\nu \bar{\psi}^c ) F_{\mu \nu}^b
           \nonumber \\
    & &  + \frac{1}{2}   (\Im {\cal F})_{ab}  D^a D^b
         + \frac{\sqrt{2}}{4} ( {\cal F}_{abc} \psi^c\lambda^a
         + \bar{{\cal F}}_{abc} \bar{\psi}^c \bar{\lambda}^a ) D^b
         + \frac{i}{4} {\cal F}_{abc} F^c \lambda^a \lambda^b
         - \frac{i}{4} \bar{{\cal F}}_{abc} \bar{F}^c \bar{\lambda}^a \bar{\lambda}^b
           \nonumber \\
    & &  - \frac{i}{8} {\cal F}_{abcd} \psi^c \psi^d \lambda^a \lambda^b
         + \frac{i}{8} \bar{{\cal F}}_{abcd} \bar{\psi}^c \bar{\psi}^d \bar{\lambda}^a \bar{\lambda}^b,
           \label{L:gauge} \\
     {\cal L}_{{\rm sup}}
    &=&    F^a \partial_a W
         - \frac{1}{2} \partial_a \partial_b W \psi^a \psi^b
         + c.c.,
           \label{L:superpotential}        
\ea
where
\ba
\fD_a = -\frac{1}{2} ({\cal F}_b f_{ac}^b \bar{\phi}^c + \bar{\cal F}_b f^b_{ac} \phi^c)
\ea
and $f_{ac}^b$ is the structure constant of $SU(N)$. 
Note that an equation of motion for $F^a$ is $F^a = -g^{ab} \overline{\partial_bW}$ + fermions. 
We also assume $\langle F^a \rangle_{{\rm tree}} = -\langle g^{ab} \overline{\partial_b W} \rangle_{{\rm tree}} =0$ at the tree level.  
At the lowest order in perturbation theory, 
there is no source which gives vev to the auxiliary field $D^0$: $\langle D^0 \rangle_{{\rm tree}} = 0$. 
The $U(N)$ gaugino is massless at the tree level while the fermionic partner of the scalar gluon receives 
 the tree level mass $m_a = m_0 = \langle g^{00} \partial_0 \partial_0 W \rangle_{{\rm tree}} $. 

\subsection{assumptions}
  While we have already stated, it is useful to recapitulate here a set of assumptions made 
  in order to address better the question of dynamical supersymmetry breaking
    within our framework. 
    
1) a general ${\cal N}=1$ supersymmetric action 
of chiral superfield $\Phi^a$ in the adjoint representation 
and the vector superfield $V^a$ with the three input functions, namely, 
the K\"{a}hler potential $K(\Phi^a, \bar{\Phi}^a)$ with its gauging, the gauge kinetic superfield $\tau_{ab}(\Phi^a)$
 that follow from the second derivatives of a generic holomorphic function ${\cal F}(\Phi^a)$, 
and the superpotential $W(\Phi^a)$.  

2) third derivatives of ${\cal F}(\Phi^a)$ at the scalar vev's are non-vanishing. 

3) the superpotential at tree level preserves ${\cal N}=1$ supersymmetry. 

4) the gauge group is $U(N)$ and the vacuum is taken to be in the unbroken phase of $U(N)$. 
It is in principle straightforward to extend this to the (partially) broken cases where $U(N)$ is broken into the product groups.
  The variational analyses we carry out in section four, however, become more complex  and
   we will not address this in this paper,  See the comment at eq. (\ref{Jintroduced})

\subsection{supercurrent}
 We give here an off-shell form of the ${\cal N}=1$ supercurrent. 
\ba
   \eta_1 {\cal S}^{(1) \mu}
   &=&   \sqrt{2} g_{ab} \eta_1 \sigma^\nu \bar{\sigma}^\mu \psi^a {\cal D}_\nu \bar{\phi}^b         
        + \sqrt{2} i g_{ab} \eta_1 \sigma^\mu \bar{\psi}^a F^b
           \nonumber \\
&&- i {\cal F}_{ab}
    \eta_1 \sigma_\nu \bar{\lambda}^a {F^{\mu \nu}}^b 
   + \frac{1}{2}  {\cal F}_{ab}  \epsilon^{\mu \nu \rho \delta} \eta_1 \sigma_\nu \bar{\lambda}^a F^{\rho \delta b}  
   - \frac{i}{2}  \bar{\cal F}_{ab}  \eta_1 \sigma^\mu \bar{\lambda}^a D^b  \nonumber \\
     && + \frac{\sqrt{2}}{4} \left( {\cal F}_{abc} \psi^c \sigma^{\nu} \bar{\sigma}^{\mu} \lambda^b
         -\bar{\cal F}_{abc}  \bar{\lambda}^c \bar{\sigma}^{\mu} \sigma^{\nu} \bar{\psi}^b \right)
            \eta_1 \sigma_\nu \bar{\lambda}^a.
                \label{susycurrent1}
  \ea       
Equations of motion for auxiliary fields are
\ba
    D^{a}
    &=&  - \frac{1}{2} g^{ab} \fD_b
         - \frac{1}{2 \sqrt{2}} g^{ab} 
           \left( {\cal F}_{bcd}\psi^d \lambda^c 
         + \bar{{\cal F}}_{bcd} \bar{\psi}^d \bar{\lambda}^c \right), 
           \nonumber \\
    F^a
    &=&  - g^{ab} \overline{\partial_{b} W}
         - \frac{i}{4} g^{ab} \left( {\cal F}_{bcd} \psi^c \psi^d 
         - \bar{{\cal F}}_{bcd} \bar{\lambda}^c \bar{\lambda}^d \right).
           \label{DFF}
\ea
 Once the $U(N)$ invariant components of the auxiliary fields, $D^0$ and $F^0$ receive  non-vanishing vev's
 together with $U(N)$ invariant scalar vev's,  the second and the fifth terms of the RHS of eq. (\ref{susycurrent1})
at these vev's clearly develop a one-body fermionic operator non-vanishing at zero momentum \cite{Oraif, SS, FI2}:
 this particular combination of $\bar{\psi}^0$ and $\bar{\lambda}^0$ creates the one-particle state 
 which is identified with the Nambu-Goldstone fermion \cite{SS}.

\subsection{original reasoning of DDSB}

In ref. \cite{imaru1}, it was shown that  the vacuum state of
 the theory, albeit being metastable, develops a non-vanishing vev of an auxiliary field $D^0$
 in the Hartree-Fock approximation. The theory, therefore, realizes the D-term dynamical supersymmetry breaking. 
The relatively simple estimate has shown  that the vacuum can be made long lived.  Let us recall a few more key aspects.

The part of the lagrangian which produces the fermion mass matrix of size $2N$ 
is
\ba
-\frac{1}{2} (\lambda^a, \psi^a) 
\left(
\begin{array}{cc}
0 & -\frac{\sqrt{2}}{4} {\cal F}_{abc} D^b \\
-\frac{\sqrt{2}}{4} {\cal F}_{abc} D^b & \partial_a \partial_c W \\
\end{array}
\right) 
\left(
\begin{array}{c}
\lambda^c \\
\psi^c \\
\end{array}
\right) + (c.c.).
\label{massterm}
\ea 
 
 It was observed that the auxiliary $D^a$ field,
  which is an order parameter of ${\cal N} =1$ supersymmetry,
    couples to the fermionic (but not bosonic) bilinears through the third prepotential derivatives:
     the non-vanishing vev of $D^0$ immediately gives a Dirac mass to the fermions.
  Eq. (\ref{DFF}) implies
  \ba
    \langle D^{0} \rangle
    =    - \frac{1}{2 \sqrt{2}} \langle g^{00} 
           \left( {\cal F}_{0cd}\psi^d \lambda^c 
         + \bar{{\cal F}}_{0cd} \bar{\psi}^d \bar{\lambda}^c \right) \rangle,
 \ea
 telling us that the condensation of the Dirac bilinear is responsible for $\langle D^{0} \rangle \neq 0$.
 
 We diagonalize the holomorphic part of the mass matrix: 
\ba
  M_{F a} \equiv
\left(
\begin{array}{cc}
0 & -\frac{\sqrt{2}}{4} \langle {\cal F}_{0aa} D^0 \rangle \\
-\frac{\sqrt{2}}{4} \langle {\cal F}_{0aa} D^0 \rangle & \langle \partial_a \partial_a W \rangle \\
\end{array}
\right). 
\label{massmatrix1}
\ea 
 Note that the non-vanishing third prepotential derivatives are ${\cal F}_{0aa}$
 where $a$ refers to the generators of the unbroken gauge group.
 By an orthogonal transformation, we obtain the two eigenvalues of eq. (\ref{massmatrix1}) for each generator, 
 which are mixed Majorana-Dirac type  :
 \ba
 \label{eigenvalue}
 {\Lambda}_{a {\bf 11}}^{(\pm)} &=&  \frac{1}{2} \langle \partial_a \partial_a W \rangle 
 \left( 1 \pm \sqrt{1 + \frac{\langle {\cal F}_{0aa}D^0 \rangle^2}{2\langle \partial_a \partial_a W\rangle^2} }\right). 
  \ea
Introducing 
\ba
 \lambda_{a{\bf 11}}^{(\pm)} \equiv \frac{1}{2}\left( 1 \pm \sqrt{1 +\Delta_{{\bf 11}}^2} \right), \quad
  \Delta_{a{\bf 11}}^2 \equiv   \frac{\langle {\cal F}_{0aa} D^0 \rangle^2}{2 \langle \partial_a \partial_a W \rangle^2},
  \label{lambdapm}
\ea
 we obtain
\ba
 | \Lambda_{a {\bf 11}}^{(\pm)} |^2 = | \langle \partial_a \partial_a W \rangle | | \lambda_{a{\bf 11}}^{(\pm )} |^2.
\ea
 
 It was also shown in ref. \cite{imaru1} that the non-vanishing $F^0$ term is induced
 by the consistency of our procedure of computation. (See also \cite{DNNS, CFK}). 
  This is because the stationary value of the scalar fields  gets shifted upon the variation (the vacuum condition). 
  The final mass formula for the $SU(N)$ fermions is to be read off from  
\ba
{\cal L}_{{\rm mass}}^{(holo)} &=& -\frac{1}{2} \langle g_{0a,a} \rangle \langle \bar{F}^0 \rangle \psi^a \psi^a 
+ \frac{i}{4} \langle {\cal F}_{0aa} \rangle \langle F^0 \rangle \lambda^a \lambda^a
-\frac{1}{2} \langle \partial_a \partial_a W \rangle \psi^a \psi^a 
+ \frac{\sqrt{2}}{4}
\langle {\cal F}_{0aa} \rangle \psi^a \lambda^a 
\langle D^0 \rangle  \nonumber \\
&\equiv& -\frac{1}{2} \sum_{a=1}^{N^2-1} \Psi(x)^{a~t} M_{a,a} \Psi^a(x), \qquad 
\Psi^a(x) = 
\left(
\begin{array}{c}
\lambda^a(x) \\
\psi^a(x) \\
\end{array}
\right).  
\ea
  We will write down the explicit form in the next subsection. See eqs. (\ref{imaru3massmatrix}), (\ref{parameters}),
  (\ref{ev1}) and (\ref{ev2}).
A main remaining point is how to establish the procedure in which the stationary values of the scalar fields,
 $D^0$ and $F^0$ perturbatively induced are determined, which we will resolve in this paper.

\subsection{quadratic part of the quantum action}
 In this subsection, we write down parts of the action with the background fields for the computation of the one-loop determinant
  in the next section. 
  The linear terms that arise upon separation into quantum fields and background fields are dropped
  as they always cancel with source terms in $\Gamma_{{\rm 1PI}}$.
  
\subsubsection{the fermionic part}
Let us extract the fermion bilinears from eqs. (\ref{L:K}), (\ref{L:gauge}) and (\ref{L:superpotential}) 
which are needed for our analysis in what follows. 
Rescaling the fermion fields so that their kinetic terms become canonical, we obtain
\ba
{\cal L}_F 
&=& - \frac{i}{2} \psi^a \sigma^\mu \partial_\mu \bar{\psi}^a
         + \frac{i}{2} (\partial_\mu \psi^a) \sigma^\mu \bar{\psi}^a 
 - \frac{i}{2} \lambda^a \sigma^\mu \partial_\mu \bar{\lambda}^a
         + \frac{i}{2} (\partial_\mu \lambda^a) \sigma^\mu \bar{\lambda}^a \nonumber \\
    & &  - \frac{1}{2} \left( g^{bb} g_{0b,\bar{b}} F^0 \right) \bar{\psi}^b \bar{\psi}^b
         - \frac{1}{2} \left( g^{bb} g_{0b,b} \bar{F}^0 \right) \psi^b \psi^b \nonumber \\
    &&     + \frac{\sqrt{2}}{4} \left( {\cal F}_{0aa} \sqrt{g^{aa}~{\rm Im}{\cal F}^{aa}} D^0 \right) \psi^a \lambda^a
         + \frac{\sqrt{2}}{4} \left( \bar{{\cal F}}_{0aa} \sqrt{g^{aa}~{\rm Im}{\cal F}^{aa}} D^0 \right) \bar{\psi}^a \bar{\lambda}^a 
         \nonumber \\
        && + \frac{i}{4} \left( {\cal F}_{0aa} g^{aa} F^0 \right) \lambda^a \lambda^a
         - \frac{i}{4} \left( \bar{{\cal F}}_{0aa} g^{aa} \bar{F}^0 \right) \bar{\lambda}^a \bar{\lambda}^a \nonumber \\
      &&   - \frac{1}{2} \left( g^{aa} \partial_a \partial_a W \right) \psi^a \psi^a 
         -\frac{1}{2} \left( g^{aa} \overline{\partial_a \partial_a W} \right) \bar{\psi}^a \bar{\psi}^a. 
\ea
Here the fermion fields $\psi^a$, $\bar{\psi}^a$, $\lambda^a$, $\bar{\lambda}^a$ are to be integrated to make a part of the effective potential, 
while the gauge kinetic function ${\cal F}_{aa}$, the K\"ahler metric $g_{aa}$ and their derivatives are functions of 
the $U(N)$ singlet $c$-number background scalar field $\varphi^0$. 
The order parameters of supersymmetry $F^0$, $\bar{F}^0$, and $D^0$ are taken as background fields as well. 

From the lagrangian ${\cal L}_F$, the holomorphic part of the mass matrix is read off as
\ba
{\cal M}_a &=&
\left(
\begin{array}{cc}
-\frac{i}{2} g^{aa} {\cal F}_{0aa} F^0, & -\frac{\sqrt{2}}{4} \sqrt{g^{aa} ({\rm Im}{\cal F})^{aa}}{\cal F}_{0aa} D^0 \\
-\frac{\sqrt{2}}{4} \sqrt{g^{aa} ({\rm Im}{\cal F})^{aa}}{\cal F}_{0aa} D^0, & g^{aa} \partial_a \partial_a W + g^{aa} g_{0a,a} \bar{F}^0 \\
\end{array}
\right) 
= \left(
\begin{array}{cc}
m_{\lambda\lambda}^a & m_{\lambda\psi}^a \\
m_{\psi \lambda}^a  
& m_{\psi\psi}^a \\
\end{array}
\right). 
\label{imaru3massmatrix}
\ea
We parametrize this matrix such that, in the case of $F^0=\bar{F}^0=0$, its form reduces to that of ref. \cite{imaru1, imaru2}.
The quantities having multiple indices such as ${\cal F}_{0aa}$ receive $U(N)$ invariant expectation values:
$\langle {\cal F}_{0aa} \rangle = \langle {\cal F}_{000} \rangle$ e.t.c.
See,  for  instance,  \cite{FIS346}.
We suppress the indices as we work with the unbroken $U(N)$ phase in this paper.
\ba
\Delta \equiv -  \frac{2m_{\lambda\psi}}{m_{\psi\psi}}, \qquad f \equiv \frac{2im_{\lambda\lambda}}{{\rm tr}{\cal M}}. 
\label{parameters}
\ea
The two eigenvalues of the holomorphic mass matrix are written as
\ba
\Lambda^{(\pm)} \equiv ({\rm tr}{\cal M}) \lambda^{(\pm)},
\label{ev1}
\ea
where
\ba
\lambda^{(\pm)} = \frac{1}{2} \left( 1 \pm \sqrt{(1+if)^2 + \left( 1+\frac{i}{2}f \right)^2 \Delta^2 } \right).
\label{ev2}
\ea
 These provide the masses for the two species of $SU(N)$ fermions once the stationary
values are determined.

\subsubsection{the bosonic part}

Likewise, we extract the bosonic quantum bilinears from eqs. (\ref{L:K}), (\ref{L:gauge}), and (\ref{L:superpotential}). 
Let 
\ba
\phi^a &=& \delta_0^a \varphi^0 + \sqrt{g^{aa}(\varphi)} \tilde{\varphi}^a, \\ 
A_\mu^a &=& \sqrt{({{\rm Im}~{\cal F}})^{aa}} \tilde{A}_\mu^a, \\
F^a &=& \sqrt{g^{aa}(\varphi)} \tilde{F}^a, \\
D^a &=& \sqrt{({\rm Im}~{\cal F})^{aa}} \tilde{D}^a
\ea
where $\varphi^0$ are the background $c$-number field while 
$\tilde{\varphi}^a$, $\tilde{A}_\mu^a$, $\tilde{F}^a$ and $\tilde{D}^a$ are the quantum scalar, vector and auxiliary fields
respectively. 

We obtain
\ba
{\cal L}_B^{(1)} &=& \partial_\mu \tilde{\varphi}^a \partial^\mu \tilde{\varphi}^{*a} 
-\frac{1}{4} \tilde{F}_{\mu \nu}^a \tilde{F}^{a \mu \nu} 
+ \tilde{F}^a \tilde{\bar{F}}^a + \frac{1}{2} \tilde{D}^a \tilde{D}^a \nonumber \\
&&+ \tilde{F}^a \left( (\sqrt{g^{aa}} \partial_a W) + (g^{aa} \partial_a \partial_a W) \tilde{\varphi}^a \right)
+ \tilde{\bar{F}}^a \left( (\sqrt{g^{aa}} \overline{\partial_a W}) + (g^{aa} \overline{\partial_a \partial_a W}) \tilde{\varphi}^{a*} \right).
\ea
We have also ignored 
$-\frac{1}{8}({\rm Re}~{\cal F})_{ab}\epsilon^{\mu\nu\rho\sigma}F_{\mu\nu}^a F_{\rho\sigma}^b$ 
as we eventually set $\varphi^a$ to be constant in our analysis and this term becomes a total derivative.

\subsection{coupling to ${\cal N}=1$ supergravity and super-Higgs mechanism}

If eq. (\ref{KtauW}) couples to ${\cal N}=1$ supergravity, 
the lagrangian is augmented to become the following one \cite{WB}:
\ba
{\cal L} &=& \int d^2 \Theta 2 {\cal E} 
\left[
\frac{3}{8} (\bar{{\cal D}}\bar{{\cal D}}-8{\cal R}) {\rm exp} 
\left\{
-\frac{1}{3} [K(\Phi, \Phi^\dag) + \Gamma(\Phi, \Phi^\dag, V)]
\right\} \right. \nonumber \\
&& \left. + \frac{1}{16g^2} \tau_{ab}(\Phi) W^{\alpha a} W^{b}_\alpha + W(\Phi)
\right] + h.c.
\label{sugra}
\ea
The fermionic part of the lagrangian relevant to the super-Higgs mechanism is given by 
\ba
e^{-1} {\cal L}_{{\rm fermionic}} &=& -i \bar{\psi}_a \bar{\sigma}^\mu \tilde{{\cal D}}_\mu \psi^a 
+ \epsilon^{\mu \nu \alpha \beta} \bar{\psi}_\mu \bar{\sigma}_\nu \tilde{{\cal D}}_\alpha \psi_\beta
-\frac{i}{2} \left[ \lambda_a \sigma^\mu \tilde{{\cal D}}_\mu \bar{\lambda}^a + \bar{\lambda}_a \bar{\sigma}^\mu \tilde{{\cal D}}_\mu \lambda^a \right] \nonumber \\
&&-\frac{i}{2 \sqrt{2}} g \partial_c \tau_{ab} D^a \psi^c \lambda^b + \frac{i}{2 \sqrt{2}} g \partial_{c^*} \tau_{ab}^* D^a \bar{\psi}^c \bar{\lambda}^a 
-\frac{1}{2} g D_a \psi_\mu \sigma^\mu \bar{\lambda}^a + \frac{1}{2} g D_a \bar{\psi}_\mu \bar{\sigma}^\mu \lambda^a \nonumber \\
&&-e^{K/2} 
\left[
W^* \psi_\mu \sigma^{\mu\nu} \psi_\nu + W \bar{\psi}_\mu \bar{\sigma}^{\mu\nu} \bar{\psi}_\nu 
+ \frac{i}{\sqrt{2}} D_a W \psi^a \sigma^\mu \bar{\psi}_\mu + \frac{i}{\sqrt{2}} D_{a^*}W^* \bar{\psi}^a \bar{\sigma}^\mu \psi_\mu \right. \nonumber \\
&& \left. + \frac{1}{2} {\cal D}_a D_b W \psi^a \psi^b + \frac{1}{2} {\cal D}_{a^*} D_{b^*} W^* \bar{\psi}^a \bar{\psi}^b \right. \nonumber \\
&& \left. -\frac{1}{4} g^{ab^*}D_{b^*} W^* \partial_a \tau_{cd} \lambda^c \lambda^d -\frac{1}{4} g^{ab^*} D_a W \partial_{b^*} \tau_{cd}^* \bar{\lambda}^c \bar{\lambda}^d
\right], 
\ea
where $e$ is the determinant of the vierbein and the covariant  derivatives of several kinds are defined as follows,
\ba
\tilde{{\cal D}}_\mu \psi_\nu &=& \partial_\mu \psi_\nu + \omega_\mu \psi_\nu + \frac{1}{4} (\partial_a K \tilde{{\cal D}}_\mu \phi^a - \partial_{a^*} K \tilde{{\cal D}}_\mu \phi^{a*})\psi_\nu 
+ \frac{i}{2}g A_\mu^a {\rm Im}F_a \psi_\nu, \\
\tilde{{\cal D}}_\mu \lambda^a &=& \partial_\mu \lambda^a+ \omega_\mu \lambda^a - g f^{abc} A_\mu^b \lambda^c 
+ \frac{1}{4} (\partial_b K \tilde{{\cal D}}_\mu \phi^b - \partial_{b^*} K \tilde{{\cal D}}_\mu \phi^{b*}) \lambda^a \nonumber \\
&& + \frac{i}{2}g A_\mu^b {\rm Im}F_b \lambda^a, \\
D_a W &=& \partial_a W + (\partial_a K) W, \\
{\cal D}_a D_bW &=& \partial_a \partial_b W + (\partial_a \partial_b K) W + 2 (\partial_a K) D_b W - (\partial_a K)(\partial_b K) W. 
\ea
Now, we focus on the gravitino mass terms to discuss super-Higgs mechanism associated with eq. (\ref{sugra}).  
\ba
e^{-1} {\cal L}_{{\rm gravitino~mass}} = - e^{K/2} W^* \psi_\mu \sigma^{\mu\nu} \psi_\nu 
+ \frac{i}{\sqrt{2}} \psi_\mu \sigma^\mu \left[ i \frac{g}{\sqrt{2}} D_a \bar{\lambda}^a + e^{K/2} D_a W^* \bar{\psi}^a \right] + h.c.
\ea
The  field redefinition of the gravitino 
\ba
\psi_\mu' = \psi_\mu + i \frac{\sqrt{2}}{6 W^* e^{K/2}} \sigma^\mu \bar{\psi}_{{\rm NG}} + \frac{\sqrt{2}}{3 W^{*2} e^K} \partial_\mu \bar{\psi}_{{\rm NG}}
\ea
eliminates the mixing terms of the gravitino with the gaugino $\lambda$ and the adjoint fermion $\psi$: 
\ba
e^{-1} {\cal L}_{{\rm gravitino~mass}} = -e^{K/2} W^* \psi_\mu' \sigma^{\mu\nu} \psi_\nu' + \frac{1}{2 W^*e^{K/2}} \bar{\psi}^2_{{\rm NG}} + h.c.
\label{gravitinomass}
\ea
where the NG fermion $\psi_{{\rm NG}}$ absorbed in the massive gravitino is read 
\ba
\bar{\psi}_{{\rm NG}} \equiv i \frac{g}{\sqrt{2}} D_a \bar{\lambda}^a + e^{K/2} D_{a^*} W^* \bar{\psi}^a. 
\ea
The eq. (\ref{gravitinomass}) tells us that the gravitino mass is given by
\ba
m_{3/2} = e^{\langle K \rangle/2} \frac{\langle W \rangle}{M_P^2}. 
\ea
Requiring the cosmological constant to be almost vanishing 
\ba
0 &\simeq& \langle V \rangle 
= \frac{g^2}{2} (D^a)^2 + e^{K} \left[ |D_a W|^2 - \frac{3}{M_P^2}|W|^2 \right], 
\ea
the gravitino mass can be expressed in terms of the vev's of the auxiliary fields
\bea
m_{3/2} \simeq e^{\langle K \rangle/2} \frac{\sqrt{|\langle  D_aW \rangle|^2 + \frac{g^2}{2}\langle D^a \rangle^2}}{\sqrt{3}M_P^2}. 
\eea

\subsection{special cases}
As is mentioned in the introduction, the theory permits a sequence of interesting limiting cases.
 If we demand  the K\"{a}hler function $K$ to be special K\"{a}hler,  $K$  are expressible in terms of ${\cal F}$
 as 
 \ba
 K =    {\rm Im}  
           {\rm Tr}~
           \bar{\Phi} 
           \frac{\partial {\cal F}(\Phi)}{\partial \Phi}, 
 \ea
   and  $g_{ab} = {\rm Im} {\cal F}_{ab}$ ${\it etc}$.
  If we further demand such that the action possesses the rigid ${\cal N}=2$ supersymmetry with one input function
 by choosing the superpotential to be a particular form, the tree vacua are shown to break ${\cal N}=2$ supersymmetry
 to ${\cal N}=1$ spontaneously \cite{APT, FIS1, FIS346, FISIMM}
\footnote{This superpotential consists of the terms referred to as the electric and magnetic Fayet-Iliopoulos terms.
This ${\cal N}=2$ FI term is very special in the sense that, by the $SU(2)_R$ rigid rotation, 
it can be represented as a part of the superpotential. 
In this way, it avoids the difficulty (see, for instance, \cite{KS} for a recent discussion) of coupling 
the system to ${\cal N}=2$ supergravity \cite{FGP, sugra}}.
We list the transformation laws for the doublet of fermions in this special case
\ba
\delta \left(
    \begin{array}{c}
    \lambda^a \\
    \psi^a
    \end{array}
    \right) 
    &=&    F_{\mu \nu}^a \sigma^{\mu \nu} 
           \left(
           \begin{array}{c}
           \eta_1 \\
           \eta_2
           \end{array}
           \right)
         - i \sqrt{2} \sigma^\mu 
           \left(
           \begin{array}{c}
           \bar{\eta}_2 \\
           - \bar{\eta}_1
           \end{array}
           \right) {\cal D}_\mu \phi^a
         + \left( \begin{array}{cc}
           i D^a & - \sqrt{2} \widetilde{F}^a \\
           \sqrt{2} F^a & i \widetilde{D}^a
           \end{array} \right)   
           \left( \begin{array}{c}
           \eta_1 \\
           \eta_2
           \end{array} \right),
\ea
where
\ba
    \widetilde{D}^{a}
    &=&  - \frac{1}{2} g^{ab} \fD_b
         + \frac{1}{2 \sqrt{2}} g^{ab} 
           \left( {\cal F}_{bcd}\psi^d \lambda^c + \bar{{\cal F}}_{bcd} 
           \bar{\psi}^d \bar{\lambda}^c \right), 
           \nonumber \\
    \widetilde{F}^a
    &=&  - \sqrt{2N} g^{ab} (e \delta_b^0 + m \bar{{\cal F}}_{0b})
         - \frac{i}{4} g^{ab} \left( {\cal F}_{bcd} \lambda^c \lambda^d 
         - \bar{{\cal F}}_{bcd} \bar{\psi}^c \bar{\psi}^d \right). 
\ea

\subsection{connection with the previous work}
We here stop shortly to address the connection of ref. \cite{imaru1} with the previous work. 
Models of dynamical supersymmetry breaking with non-vanishing F- and D-terms 
have been previously proposed: they are, for instance, the 3-2 model \cite{ADS3} 
and the 4-1 model in \cite{DNNS}.\footnote{Application of these models to the mediation mechanism, see for example \cite{CFK, CR, Caracciolo}.} 
In these models, supersymmetry is unbroken at the tree level and 
  is broken by the non-vanishing vev of the F-term through instanton generated superpotentials. 
Non-vanishing vev of the D-term  is also induced, but is much smaller than that of the F-term. 

In our mechanism, supersymmetry is unbroken at the tree level, 
 and is broken in a self-consistent Hartree-Fock approximation of  the NJL type
  that produces a non-vanishing vev for the D-term.
 A non-vanishing vev for the F-term is induced in our Hartree-Fock vacuum that shifts the tree vacuum
  and we explore the region of the parameter space in which F-term vev is treated perturbatively.
 
  We should mention that  the way in which the two kinds of gauginos (or the gaugino and the adjoint matter fermion)
   receive masses    is  an extension of that proposed in \cite{FNW}: 
 the pure Dirac-type gaugino mass is generated in \cite{FNW} \footnote{Attention  has been paid to Dirac gaugino
in many papers \cite{CR, Caracciolo, Fayet, PS, HR, NRSU, KPW, Benaklietal, ABFP, DMRM, KM, YMFJK, FKMT, Fok, Unwin, MNS}.}, 
while the mixed Majorana-Dirac type gaugino masse is generated in our case, the Majorana part being given by the
 second derivative of the superpotential. 
In \cite{FNW}, the dynamical origin of non-vanishing D-term vev was not addressed.

 As for the application to dynamical chiral symmetry breaking,  
 a supersymmetric NJL type model has been considered \cite{BL, CLB, MR, FJK}. 
Chiral symmetry is not spontaneously broken in a supersymmetric case. 
Even in softly broken supersymmetric theories, 
the chiral symmetry broken phases are degenerate with the chirally symmetric ones. 
Thus, in supersymmetric theories, the phase with broken chiral symmetry is no longer the energetically preferred ground state.

\section{The effective potential in the Hartree-Fock approximation}

The goal of this section is to determine the effective potential to the leading order in the Hartree-Fock approximation. 
We will regulate one-loop integral by the dimensional reduction \cite{Siegel}. 
We prepare a supersymmetric counterterm, setting the normalization condition. 
We make brief comments on  regularization and  subtraction schemes in the end of section 4. 
 We also change the notation for  expectation values in general from $\langle ... \rangle$ to $..._*$
  as our main thrust of this paper is the determination of the stationary values from the variational analysis.

\subsection{the point of the approximation}

In the Hartree-Fock approximation, one begins with considering the situation 
where one-loop corrections in the original expansion in $\hbar$ become large 
and are comparable to the tree contribution. 
The optimal configuration of the effective potential to this order is found 
by matching the tree against one-loop, varying with respect to the auxiliary fields. 
In this section, we start the analysis of this kind for our effective potential. 
There are three constant background fields as arguments of the effective potential: 
$\varphi \equiv \varphi^0~({\rm complex})$, $U(N)$ invariant background scalar, 
$D \equiv D^0~({\rm real})$ and $F \equiv F^0~({\rm complex})$. 
The latter two are the order parameters of ${\cal N}=1$ supersymmetry. 

We vary our effective potential with respect to all these constant fields and examine the stationary conditions. 
We  also examine a second derivative at the stationary point along the constraints of the auxiliary fields 
to understand better  the Hartree-Fock corrected mass of the scalar gluons. 
Let us denote our effective potential by $V$. 
It consists of three parts: 
\ba
V = V^{{\rm tree}} + V_{{\rm c.t.}} + V_{{\rm 1-loop}}. 
\ea
The first term is the tree contributions, the second one is the counterterm and the last one is the one-loop contributions. 
After the elimination of the auxiliary fields, 
the effective potential is referred to as the scalar potential so as to be distinguished from the original $V$. 

\subsection{the tree part}

To begin with, let us write down the tree part and find a parametrization by two complex and one real parameters. 
We also introduce simplifying notation $g_{00}(\varphi, \bar{\varphi}) \equiv g(\varphi, \bar{\varphi}), 
({\rm Im}~{\cal F}(\varphi))_{00} \equiv {\rm Im}~{\cal F}''(\varphi), \partial_0 W(\varphi) = W'(\varphi), g_{00,0} \equiv \partial g,$ etc. 
\ba
V^{{\rm tree}}(D, F, \bar{F}, \varphi, \bar{\varphi}) =-gF\bar{F} -\frac{1}{2} ({\rm Im}{\cal F}'') D^2 - F W' - \bar{F}\bar{W'}.  
\label{treepart}
\ea
As a warm up, let us determine the vacuum configuration by a set of stationary conditions at the tree level:
\ba
&&\frac{\partial V^{{\rm tree}}}{\partial D} = 0, \label{dflattree} \\
&&\frac{\partial V^{{\rm tree}}}{\partial F} = 0, {\rm as~well~as~its~complex~conjugate}, \label{fflattree} \\
&&\frac{\partial V^{{\rm tree}}}{\partial \varphi} = 0, {\rm as~well~as~its~complex~conjugate}.  \label{phistattree}
\ea
Eq. (\ref{dflattree}) determines the stationary value of $D$:
\ba
D=0 \equiv D_*, 
\ea
while from eq. (\ref{fflattree}), we obtain 
\ba
F=-g^{-1}(\varphi, \bar{\varphi}) \bar{W}'(\bar{\varphi}) \equiv F_*(\varphi, \bar{\varphi}).
\ea
Eq. (\ref{phistattree}) together with these two gives
\ba
W'(\varphi_*) = 0~{\rm and~therefore}~F_*(\varphi, \bar{\varphi})=0,
\ea
as well as
\ba
V_{{\rm scalar}}^{{\rm tree}} (\varphi, \bar{\varphi}) \equiv V^{{\rm  tree}} (\varphi, \bar{\varphi}, D_*=0, F=F_*(\varphi, \bar{\varphi}), 
\bar{F}=\overline{F_*(\varphi, \bar{\varphi})}) = g^{-1}(\varphi, \bar{\varphi}) |W'(\varphi)|^2. 
\ea
The negative coefficients of the RHS of eq. (\ref{treepart}) imply that both $D$ and $F$ profiles of the potential
have a maximum for a given $\varphi$. These signs are, of course, the right signs for the stability of the scalar potential
 as is clear by completing the square. This is a trivial comment to make here but will become less trivial later.
The mass of the scalar gluons at tree level $|m_{s*}|^2$ is read off from the second derivative at the stationary point:
\ba
\left. 
\frac{\partial^2 V^{{\rm tree}}(\varphi, \bar{\varphi})}{\partial \varphi \partial \bar{\varphi}}
\right|_{\varphi_*, \bar{\varphi}_*} 
&=& g^{-1}(\varphi_*, \bar{\varphi}_*) \left| W''(\varphi_*)\right|^2, \\
m_s(\varphi, \bar{\varphi}) &\equiv& g^{-1}(\varphi, \bar{\varphi}) W''(\varphi), \label{sgluonmass} \\
m_{s*} &=& m_s (\varphi_*, \bar{\varphi}_*).  
\ea
As we have already introduced in eq. (\ref{parameters}), $\Delta$ and $r$ are defined by 
\ba
\Delta \equiv -2 \frac{m_{\lambda \psi}}{m_{\psi \psi}} = \frac{\sqrt{2}}{2} 
\frac{\sqrt{g^{-1}({\rm Im}{\cal F}'')^{-1}} {\cal F}'''}{g^{-1}W''+g^{-1}\partial g \bar{F}}~D
\equiv r(\varphi, \bar{\varphi}, F, \bar{F}) D.
\ea
Recall that we have suppressed the indices, invoking the $U(N)$ invariance of the expectation values. 
Also
\ba
f_3 \equiv \frac{g^{-1}{\cal F}''' F}{g^{-1}W'' + g^{-1} \partial g \bar{F}}, 
\label{f3}
\ea
where $f_3$ differs from $f$ in eq. (\ref{parameters}) by
\ba
(g^{-1}W'' + g^{-1} \partial g \bar{F}) f_3 = \left( g^{-1}W'' + g^{-1} \partial g \bar{F} -\frac{i}{2}g^{-1}{\cal F}'''F \right)f. 
\ea 
We obtain 
\ba
F=\frac{m_s}{g^{-1}{\cal F}'''} \varepsilon, \qquad \bar{F}=\frac{\bar{m}_s}{g^{-1} \bar{{\cal F}}'''} \varepsilon, \qquad 
\varepsilon = \frac{f_3 + \frac{\bar{m}_s}{m_s} \frac{g^{-1}\partial g}{g^{-1}\bar{{\cal F}}'''}  |f_3|^2}{1 - \left| \frac{g^{-1}\partial g f_3}{g^{-1}{\cal F}'''} \right|^2}. 
\ea
While we will not make exploit in this paper, $V^{{\rm tree}}$ can be written as a function of $\varphi$ complex, $|\Delta|$ real, $f_3$ complex:
\ba
V^{{\rm tree}} = -  \left| \bar{m}_s + \frac{g^{-1} \bar{\partial} g}{g^{-1}{\cal F}'''} m_s \varepsilon \right|^2 |{\cal F}'''|^{-2} g^2
\left(
g^{-1}({\rm Im}{\cal F}'') |\Delta|^2 + g |f_3|^2
\right)
- \frac{m_s}{g^{-1} {\cal F}'''} \varepsilon W'
- \frac{\bar{m}_s}{g^{-1} \overline{{\cal F}}'''} \bar{\varepsilon} \bar{W}' \nonumber \\
\ea
where $m_s, \varepsilon, g, {\cal F}$ (and their derivatives) are the functions listed above
and undergo the variations to be carried out in the subsequent subsections. 
We also see that the mass scales of the problem are set by $m_{s*}$, 
the scalar gluon mass  and  $g^{-1} \overline{{\cal F}}'''_*$, the third prepotential derivative, (and $g^{-1} \partial g$), 
once the stationary value of the scalar is determined.

\subsection{treatment of UV infinity}

In the NJL theory \cite{NJL1, NJL2}, there is only one coupling constant carrying dimension $-1$ and 
the dimensionless quantity is naturally formed by combining it with the relativistic cutoff, 
which is interpreted as the onset of UV physics. 
In the theory under our concern, 
UV physics is specified by the three input functions, $K, {\cal F}, W$ 
and the UV scales and infinities reside in some of the coefficients. 
Our supersymmetric counterterm \cite{imaru1, imaru2} is
\ba
V_{{\rm c.t.}} = -\frac{1}{2} {\rm Im} \int d^2\theta \Lambda {\cal W}^{0\alpha} {\cal W}_{0\alpha} 
= -\frac{1}{2} ({\rm Im} \Lambda) D^2. 
\ea
It is a counterterm associated with ${\rm Im}{\cal F}''$. 
We set up a renormalization condition
\ba
\left. \frac{1}{N^2} \frac{\partial^2 V}{(\partial D)^2}
 \right|_{D=0, \varphi = \varphi_*, \bar{\varphi} = \bar{\varphi}_* } =2c, 
\label{rencond}
\ea 
and relate (or transmute) the original infinity of the dimensional reduction scheme with that of ${\rm Im}{\cal F}''$. 
 We have indicated that this condition is set up at $D=0$ and the stationary point of the scalar which we will determine.
We stress again that the entire scheme is supersymmetric. 

\subsection{the one-loop part}

The entire contribution of all particles in the theory to $i \cdot$ (the 1PI to one-loop) $\equiv i \Gamma_{1-loop}$ 
is easy to compute, knowing (\ref{ev1}), (\ref{ev2}) and (\ref{sgluonmass}).
It is given by
\ba
i \Gamma_{{\rm 1-loop}} = 
\left(
\int d^4x
\right)
\sum_a \int \frac{d^4k}{(2\pi)^4} \ln 
\left(
\frac{(|\Lambda_a^{(+)}|^2 - k^2 - i \varepsilon)(|\Lambda_a^{(-)}|^2 - k^2 - i \varepsilon)}{(|m_{s,a}|^2 - k^2 - i \varepsilon)(-k^2 - i \varepsilon)}
\right). 
\label{1loopeffaction}
\ea
In the unbroken $U(N)$ phase, 
it is legitimate to replace $\displaystyle  \sum_a$ by $N^2$ and drop the index $a$ 
as we have said before.\footnote{In those cases where the $U(N)$ is broken to the product group $\prod_{\alpha =1}^n U(N_{\alpha})$, 
we need not only replace $\displaystyle  \sum_a  \cdots$  by $\displaystyle  \sum_{\alpha}  \cdots_{\alpha}$ 
but also must treat the ${\cal N} =1$ multiplet of the broken generators that receives the mass by the Higgs mechanism \cite{FIS346}.} 
We obtain 
\ba
V_{{\rm 1-loop}} &\equiv& (-i) \frac{1}{(\int d^4x)} \Gamma_{{\rm 1-loop}} \\
&=& -N^2 \left| {\rm tr}{\cal M} \right|^4 \int \frac{d^4l^\mu}{(2\pi)^4 i} 
\ln 
\left(
\frac{(|\lambda^{(+)}|^2 - l^2 - i \varepsilon)(|\lambda^{(-)}|^2 - l^2 - i \varepsilon)}{(\left| \frac{m_{s}}{{\rm tr}{\cal M}} \right|^2 - l^2 - i \varepsilon)(-l^2 - i \varepsilon)}
\right) \nonumber \\
&\equiv& N^2 |{\rm tr}{\cal M}|^4 J. 
\label{Jintroduced}
\ea
Note that $|m_s|^2$, whose stationary value give the tree mass squared of the scalar gluon, 
differ from $|{\rm tr}{\cal M}|^2$: 
\ba
\left| {\rm tr}{\cal M} \right|^2 = \left| m_s -\frac{i}{2} (g^{-1}{\cal F}''') F + (g^{-1}\partial g) \bar{F} \right|^2. 
\ea
To evaluate the integral in d-dimensions, we just quote
\ba
I(x^2) &\equiv& -\int \frac{d^4l^\mu}{(2\pi)^4 i} \log (x^2 - l^2 - i \varepsilon), \\
I(x^2)-I(0) &=& \frac{1}{32\pi^2} \left[ A(\varepsilon, \gamma) (x^2)^2 -(x^2)^2 \log (x^2) \right]
\ea
where
\ba
A(\varepsilon, \gamma) = \frac{1}{2} -\gamma +\frac{1}{\varepsilon}, \qquad \varepsilon = 2-\frac{d}{2}. 
\ea
We obtain
\ba
V_{{\rm 1-loop}} &=& \frac{N^2|{\rm tr}{\cal M}|^4}{32\pi^2} 
\left[
A(\varepsilon, \gamma) 
\left(
|\lambda^{(+)}|^4 + |\lambda^{(-)}|^4 - \left|\frac{m_s}{{\rm tr}{\cal M}} \right|^4
\right) \right. \nonumber \\
&& \left. 
-|\lambda^{(+)}|^4\log |\lambda^{(+)}|^2 
-|\lambda^{(-)}|^4\log |\lambda^{(-)}|^2
+\left| \frac{m_s}{{\rm tr}{\cal M}} \right|^4 \log \left| \frac{m_s}{{\rm tr}{\cal M}} \right|^4
\right].
\label{1loopeffpot}
\ea
This again depends upon $\Delta$, $f$ and $\varphi$.

\section{Stationary conditions and gap equation}
\subsection{variational analyses}

Now we turn to our variational problem. 
It is stated as in the tree case as
\ba
&&\frac{\partial V}{\partial D} = 0, \label{Dflat} \\
&&\frac{\partial V}{\partial F} = 0~{\rm and~its~complex~conjugate}, \label{Fflat}\\
&&\frac{\partial V}{\partial \varphi} = 0~{\rm and~its~complex~conjugate}. \label{phiflat}
\ea
We will regard the solution to be obtained by considering eqs. (\ref{Dflat}) and (\ref{phiflat}) first and 
solving $D$ and $\varphi$ for $F$ and $\bar{F}$:
\ba
D=D_*(F, \bar{F}), \quad \varphi = \varphi_*(F, \bar{F}), \quad \bar{\varphi} = \bar{\varphi}_*(F, \bar{F}).
\ea
Eq. (\ref{Fflat}) is then
\ba
\left. 
\frac{\partial V(D=D_*(F, \bar{F}), \varphi = \varphi_*(F, \bar{F}), \bar{\varphi} = \bar{\varphi}_*(F, \bar{F}), F, \bar{F})}{\partial F}
\right|_{D, \varphi, \bar{\varphi}, \bar{F}~{\rm fixed}} =0
\ea
and its complex conjugate. 
These will determine $F=F_*, \bar{F}=\bar{F}_*$. 

In this paper, we are going to work in the region
where the strength $||F_*||$ is small and can be treated perturbatively. 
This means that, in the leading order, the problem posed by eq. (\ref{Dflat}) and eq. (\ref{phiflat}) becomes
\ba
&&\frac{\partial V(D, \varphi, \bar{\varphi}, F=0, \bar{F}=0)}{\partial D} = 0, \label{Dflatleading} \\
&&\frac{\partial V(D, \varphi, \bar{\varphi}, F=0, \bar{F}=0)}{\partial \varphi} =  \frac{\partial V(D, \varphi, \bar{\varphi}, F=0, \bar{F}=0)}{\partial \bar{\varphi}} = 0
\label{phiflatleading}
\ea
and this problem does not involve the tree potential eq. (\ref{treepart}) except the last $D^2$ term,
 as $F$, and $\bar{F}$ are set zero. 
Eq. (\ref{Dflatleading}) is nothing but the gap equation given in \cite{imaru1, imaru2}, 
while eq. (\ref{phiflatleading}) is the stationary conditions for the scalar.  
This is the variational problem which is analyzed in this paper. 
A set of stationary values $(D_*, \varphi_*, \bar{\varphi}_*)$ is determined as the solution.

\subsection{the analysis in the region $F_* \approx 0$}

Let us first determine $V(D, \varphi, \bar{\varphi}, F=0, \bar{F}=0)$ explicitly. 
We need to solve the normalization condition. 
\ba
2c N^2 = \left. \frac{\partial^2 V}{(\partial D)^2} \right|_{D=0,_*} 
= -({\rm Im} {\cal F}''_*) - ({\rm Im} \Lambda) + N^2 |{\rm tr}{\cal M}|^4 
\left. \frac{\partial^2 J}{(\partial D)^2} \right|_{D=0},
\ea
where $J$ has been introduced in eq.(\ref{Jintroduced}).
At $F, \bar{F} \to 0$, 
\ba
\Delta &\to& \Delta_0 \equiv r_0(\varphi, \bar{\varphi}) D, \quad {\rm where} \quad
r_0 = \frac{\sqrt{2}}{2} \frac{\sqrt{g^{-1}({\rm Im}{\cal F''})^{-1}}{\cal F}'''}{g^{-1}W''}, \\
\lambda^{(\pm)} &\to& \lambda^{(\pm)}_0 = \frac{1}{2} \left( 1\pm \sqrt{1 + \Delta_0^2} \right), 
\ea
where
\ba 
\lambda_0^{(+)} + \lambda_0^{(-)} =1, \quad \lambda_0^{(+)} \lambda_0^{(-)} = -\frac{1}{4}\Delta_0^2, \quad 
\lambda_0^{(+)} -\lambda_0^{(-)} = \sqrt{1+\Delta_0^2}, 
\ea
\ba 
&&\frac{m_s}{{\rm tr}{\cal M}} \to 1, \\
&&J \to J_0 \equiv \frac{1}{32\pi^2} 
\left[
A(\varepsilon, \gamma) \left\{ \frac{1}{2} \left( 1+\frac{1}{2}\Delta_0^2 \right)\left( 1+\frac{1}{2} \bar{\Delta}_0^2\right) 
+ \frac{1}{2}\sqrt{1+ \Delta_0^2} \sqrt{1 + \bar{\Delta}_0^2} -1
\right\} \right. \nonumber \\
&& \left. \hspace*{3.5cm} -|\lambda_0^{(+)}|^4 \log |\lambda_0^{(+)}|^2 - |\lambda_0^{(-)}|^4 \log |\lambda_0^{(-)}|^2
\right],
\ea
essentially reducing the situation to that of refs. \cite{imaru1, imaru2}. 

Note, however, that $r$ and $\Delta~({\rm or}~r_0, \Delta_0)$ are complex in general except those special cases 
which include the case of the rigid ${\cal N}=2$ supersymmetry partially broken to ${\cal N}=1$ at the tree vacua. 
For $|\Delta_0| \ll 1$, 
\ba
J_0 \approx \frac{1}{32\pi^2} 
\left[
A(\varepsilon, \gamma) \frac{1}{2}(\Delta_0^2 + \bar{\Delta}_0^2) -\frac{1}{4} (\Delta_0^2 + \bar{\Delta}_0^2) + {\cal O}(|\Delta_0|^{4-\varepsilon})
\right]
\ea
We solve the normalization condition for the number $A$ to obtain
\ba
A &=& \frac{1}{2} + \frac{32\pi^2}{|m_{s*}|^4(r_{0*}^2+\bar{r}_{0*}^2)} \left (2c + \frac{{\rm Im}{\cal F}''_*}{N^2} + \frac{{\rm Im}\Lambda}{N^2} \right) 
\equiv \tilde{A}(c, \Lambda, \varphi_*, \bar{\varphi}_*). 
\ea
We obtain
\ba
V_0 &=& V(D, \varphi, \bar{\varphi}, F=0, \bar{F}=0) \nonumber \\
&=& -\frac{1}{2} {\rm Im}{\cal F}'' D^2 -\frac{1}{2} ({\rm Im}\Lambda) D^2 \nonumber \\
&&+ \frac{N^2|m_s|^4}{32\pi^2} 
\left[
\tilde{A}(c, \Lambda, \varphi_*, \bar{\varphi}_*) 
\left\{
\frac{1}{2} \left( 1+ \frac{1}{2} \Delta_0^2 \right) \left( 1+ \frac{1}{2} \bar{\Delta}_0^2 \right) + \frac{1}{2} \sqrt{1+\Delta_0^2} \sqrt{1+\bar{\Delta}_0^2} -1
\right\} \right. \nonumber \\
&& \left. -|\lambda_0^{(+)}|^4 \log |\lambda_0^{(+)}|^2 - |\lambda_0^{(-)}|^4 \log |\lambda_0^{(-)}|^2
\right]. 
\label{renpotential}
\ea
After some calculation, this is found to be expressible as
\ba
\frac{V_0}{N^2 |m_s|^4} &=& 
\left(
\frac{1}{64\pi^2} + \tilde{c} - \tilde{\delta}(\varphi, \bar{\varphi})
\right)
\left( \frac{\Delta_0 + \bar{\Delta}_0}{2} \right)^2 
+ \frac{1}{32\pi^2} \tilde{A} \left( \frac{1}{8} |\Delta_0|^4 + f(\Delta_0, \bar{\Delta}_0) \right) \nonumber \\
&&-\frac{1}{32\pi^2} \left( |\lambda_0^{(+)}|^4 \log |\lambda_0^{(+)}|^2 + |\lambda_0^{(-)}|^4 \log |\lambda_0^{(-)}|^2 \right), 
\label{renpotential2}
\ea
where
\ba
\tilde{c} &=& \frac{c}{|m_{s*}|^4 \left( \frac{r_{0*}^2 + \bar{r}_{0*}^2}{2} \right)}, \\
\tilde{\delta}(\varphi, \bar{\varphi}) &=& \frac{1}{2} 
\left(
\frac{ \frac{{\rm Im}{\cal F}_*''}{N^2} + \frac{{\rm Im}\Lambda}{N^2} }{ \frac{ (r_{0*}^2 + \bar{r}_{0*}^2) }{2} |m_{s*}|^4} 
\right)
\left[
\frac{\frac{{\rm Im}{\cal F}''/N^2 + {\rm Im} \Lambda/N^2}{{\rm Im}{\cal F}_*''/N^2 
+ {\rm Im}\Lambda/N^2} }{\frac{|m_s|^4}{|m_{s*}|^4} \frac{\left(\frac{r_0+\bar{r}_0}{2} \right)^2}{\left( \frac{r_{0*}^2 + \bar{r}_{0*}^2}{2} \right) } } 
- 1
\right], 
\\
f(\Delta_0, \bar{\Delta}_0) &=& \frac{1}{2} \left( \sqrt{1+\Delta_0^2} \sqrt{1+\bar{\Delta}_0^2} - |\Delta_0|^2 -1 \right).
\ea
Note that 
\ba
\tilde{\delta}_* \equiv \tilde{\delta}(\varphi_*, \bar{\varphi}_*) \ne 0,
\ea
and
\ba
\left| f(\Delta_0, \bar{\Delta}_0) \right| \le {\rm const} \quad {\rm for}~\left| \Delta_0 \right| \gg 1. 
\ea

If $r_0$ (and $\Delta_0$) is real, this is rewritten as
\ba
\frac{V_0}{N^2|m_s|^4} = \left( \left(c' + \frac{1}{64\pi^2} \right) -\delta \right) \Delta_0^2 
+ \frac{1}{32\pi^2} 
\left[
\frac{\tilde{A}}{8} \Delta_0^4 
- {\lambda_0^{(+)}}^4 \log {\lambda_0^{(+)}}^2 - {\lambda_0^{(-)}}^4 \log {\lambda_0^{(-)}}^2
\right],  \nonumber \\
\label{realpotential}
\ea
where $c' \equiv \frac{c}{r_{0*}^2 |m_{s*}|^4}$ is the rescaled number, and 
\ba
\delta(\varphi, \bar{\varphi}) \equiv \frac{1}{2} 
\left(
\frac{\frac{{\rm Im}{\cal F}''_*}{N^2} + \frac{{\rm Im}\Lambda}{N^2}}{r_{0*}^2 |m_{s*}|^4}
\right)
\left[
\frac{\frac{{\rm Im}{\cal F}''/N^2+ {\rm Im}\Lambda/N^2}{{\rm Im}{\cal F}''_*/N^2 + {\rm Im}\Lambda/N^2}}{\frac{r_0^2|m_s|^4}{r_{0*}^2 |m_{s*}|^4}}
-1
\right]
\label{delta}
\ea
and $m_s(\varphi, \bar{\varphi})=g^{-1}W''$ are  the functions of $\varphi, \bar{\varphi}$. 
Clearly, there are two scales in our current problem $|r_{0*}|^{-1/2}$ and $|m_{s*}|$, 
which are controlled by the second superpotential derivative and the third prepotential derivative at the stationary value $\varphi_*$. 

Let us turn to the gap equation 
\ba
\left. \frac{\partial V_0}{\partial D} \right|_{\varphi, \bar{\varphi}} = 0. 
\ea
For eq. (\ref{renpotential2}), scaling out $|r_0|^2$, we obtain
\ba
0 &=& D
\left[
\left( \frac{1}{64\pi^2} + \tilde{c} - \tilde{\delta} \right) (1 + \cos 2 \theta) 
+ \frac{\tilde{A}}{32\pi^2} 
\left\{
\frac{1}{2}|\Delta_0|^2 - (1 - \cos 2 (\theta - \theta'))
\right\} \right. \nonumber \\
&& \left. 
-\frac{1}{32\pi^2} 
\left\{
\left( 2 \log |\lambda_0^{(+)}|^2 + 1 \right) \frac{1}{2} 
\left(
\frac{e^{2i\theta}\bar{\lambda}_0^{(+)}}{\sqrt{1+\Delta_0^2}} + \frac{e^{-2i\theta}\lambda_0^{(+)}}{\sqrt{1+\bar{\Delta}_0^2}}
\right) |\lambda_0^{(+)}|^2 
\right. \right. \nonumber \\
&& \left. \left. 
- \left( 2 \log |\lambda_0^{(-)}|^2 + 1 \right) \frac{1}{2} 
\left(
\frac{e^{2i\theta}\bar{\lambda}_0^{(-)}}{\sqrt{1+\Delta_0^2}} + \frac{e^{-2i\theta}\lambda_0^{(-)}}{\sqrt{1+\bar{\Delta}_0^2}}
\right) |\lambda_0^{(-)}|^2
\right\}
\right], 
\label{gap1}
\ea
where 
\ba
\Delta_0 = |\Delta_0| e^{i\theta}, \quad r_0 = |r_0| e^{\i\theta}, \quad \tan 2\theta' = \frac{|\Delta_0|^2 \sin 2\theta}{1+ |\Delta_0|^2 \cos 2\theta}. 
\ea
Note that $|1-\cos 2(\theta-\theta')| \to 0$ in the region $\theta \sim 0$ or $|\Delta_0| \gg 1$. 

On the other hand, for eq. (\ref{realpotential}) with $\Delta_0$ being real, $N^2 |m_s|^4$ is scaled out and 
it is simply given by the $\Delta_0$ derivative:
\ba
0 &=& \Delta_0 
\left[
2 \left( \left( c'+ \frac{1}{64\pi^2} \right) - \delta \right) 
+ \frac{1}{32 \pi^2}
\left\{
\frac{\tilde{A}}{2} \Delta_0^2 
+ \frac{1}{\Delta_0} \frac{d}{d \Delta_0} 
\left(
- \lambda_0^{(+)4} \log \lambda_0^{(+)2} - \lambda_0^{(-)4} \log \lambda_0^{(-)2}
\right)
\right\}
\right] \nonumber \\
&=&
\Delta_0 
\left[
2 \left( 
\left( c' + \frac{1}{64 \pi^2} \right) - \delta 
\right) \right. \nonumber \\ 
&& \left. 
+ \frac{1}{32\pi^2}
\left\{
\frac{\tilde{A}}{2} \Delta_0^2 
-\frac{1}{\sqrt{1 + \Delta_0^2}} 
\left(
\lambda_0^{(+)3} \left( 2 \log \lambda_0^{(+)2} + 1 \right)
- \lambda_0^{(-)3} \left( 2 \log \lambda_0^{(-)2} + 1 \right)
\right)
\right\}
\right],
\label{gap2}
\ea
which is our original gap equation.\footnote{We have introduced $\delta(\varphi, \bar{\varphi})$ 
such that its stationary value $\delta(\varphi_*, \bar{\varphi}_*)=0$, which can therefore be ignored in
 analyzing eq. (\ref{gap2}).}
In both cases, the solutions are given by the extremum of the potential $V_0(D, \varphi, \bar{\varphi})$ in its $D$ profile. 
We stress again that the $D$ profile is not a direct stability criterion of the vacua, 
which is to be discussed with regard to the scalar potential $V_0(D_*(\varphi, \bar{\varphi}), \varphi, \bar{\varphi})$.

We next examine $\left. \frac{\partial V_0}{\partial \varphi} \right|_{D, \bar{\varphi}}=0$ and its complex conjugate. 
For eq. (\ref{renpotential2}), we obtain
\ba
2 \frac{\partial}{\partial \varphi} (\ln |m_s|^2) \frac{V_0}{N^2 |m_s|^4} &=& \left( \frac{\partial \tilde{\delta}}{\partial \varphi} \right) 
\left( \frac{\Delta_0 + \bar{\Delta}_0}{2} \right)^2 
-D \left[ \left. \frac{\partial \ln r_0}{\partial \varphi} \right|_{\bar{\varphi}} + \left. \frac{\partial \ln \bar{r}_0}{\partial \bar{\varphi}} \right|_{\varphi} \right]
\frac{\partial}{\partial D} \left( \frac{V_0}{N^2 |m_s|^4} \right) \nonumber \\
&& -D \hat{{\cal P}} \left( \frac{V_0}{N^2 |m_s|^4} \right),~{\rm and~its~complex~conjugate}
\label{phiflat1}
\ea
where
\ba
\hat{{\cal P}} = i \left( \frac{\partial \theta}{\partial \varphi} \right) 
\left(
r_0 \frac{\partial}{\partial \Delta_0} - \bar{r}_0 \frac{\partial}{\partial \bar{\Delta}_0}
\right). 
\ea
The second term of the RHS of eq. (\ref{phiflat1}) is proportional to the gap equation eq. (\ref{gap1}). 
As for the third term,  after some calculation, we obtain
\ba
\frac{-\hat{{\cal P}}\left( \frac{V_0}{N^2 |m_s|^4} \right)}{\left(\frac{\partial \theta}{\partial \varphi} \right)|\Delta||r_0|} 
&=& \left( \frac{1}{64\pi^2} + \tilde{c} - \tilde{\delta} \right) \sin 2 \theta 
+ \frac{1}{32\pi^2} \tilde{A} \sin 2(\theta-\theta') \nonumber \\
&&- \frac{1}{32\pi^2} \frac{1}{2} \left( \frac{\sin(2\theta-\theta')}{|1+\Delta_0^2|^{1/2}} + \sin 2 (\theta - \theta') \right) 
|\lambda_0^{(+)}|^2 \left( 2 \log|\lambda_0^{(+)}|^2 +1 \right) \nonumber \\
&&+ \frac{1}{32\pi^2} \frac{1}{2} \left( \frac{\sin(2\theta-\theta')}{|1+\Delta_0^2|^{1/2}} - \sin 2 (\theta - \theta') \right) 
|\lambda_0^{(-)}|^2 \left( 2 \log |\lambda_0^{(-)}|^2 +1 \right) \nonumber \\
&&\equiv C(\theta, |\Delta_0|). 
\label{phiflat11}
\ea
In the RHS of eqs. (\ref{phiflat1}) and (\ref{phiflat11}), 
we have regarded $\Delta_0, \bar{\Delta}_0, \varphi$ and $\bar{\varphi}$ as independent variables. 

For eq. (\ref{realpotential}), with $\Delta_0$ real, we obtain
\ba
2 \partial(\ln |m_s|^2) \frac{V_0}{N^2|m_s|^4} = \left( \frac{\partial \delta}{\partial \varphi} \right) \Delta_0^2 
-\frac{\partial \Delta_0}{\partial \varphi} \frac{\partial}{\partial \Delta_0} \left(\frac{V_0}{N^2 |m_s|^4} \right)
\label{phiflat2}
\ea
and its complex conjugate. 
Here in the last term of the RHS, we have regarded $\Delta_0, \varphi, \bar{\varphi}$ as independent variables. 
\begin{figure}[htbp]
 \begin{center}
  \includegraphics[width=60mm]{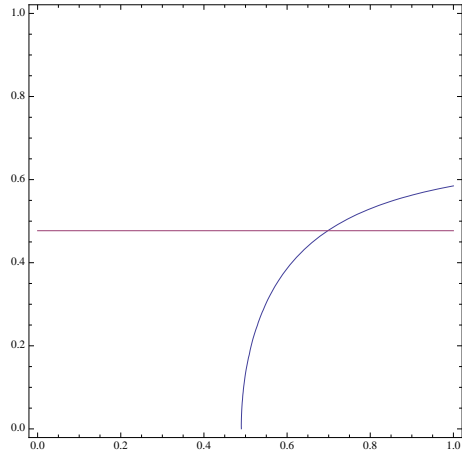}
 \end{center}
 \caption{The schematic picture of the intersection of the two curves which represent 
 the solution to the gap equation (the red one) 
 and the $\varphi$ flat condition (the blue one). The horizontal axis is denoted by $\varphi/M$ and 
the vertical one by $\Delta_0$. 
 The values at the stationary point ($\Delta_{0*}, \varphi_*=\bar{\varphi}_*$) are
  read off from the intersection point. }
 \label{intersection}
\end{figure}

Finally the stationary values ($D_*, \varphi_*, \bar{\varphi}_*$) are determined by eqs. (\ref{gap1}) and 
(\ref{phiflat1}) or by eqs. (\ref{gap2}) and (\ref{phiflat2}).  
Let us discuss the latter case first. As the second term of the RHS in eq. (\ref{phiflat2}) is nothing but
 the gap equation eq. (\ref{gap2}), 
eq. (\ref{phiflat2}) can be safely replaced by 
\ba
\frac{V_0}{N^2|m_s|^4} = \frac{\frac{\partial \delta}{\partial \varphi}}{2 \partial (\ln |m_s|^2)} \Delta_0^2,
\label{7}
\ea
$\varphi$ being real. 
The solution to eq. (\ref{7}) in the $\Delta_0$ profile is determined as the point of intersection of the potential 
with the quadratic term having $\varphi=\bar{\varphi}$ dependent coefficients. 
Actually, it is a real curve in the full ($\Delta_0, \varphi=\bar{\varphi}$) plane. 
Likewise, the solution to the gap equation eq. (\ref{gap2}), the condition of $\Delta_0$ extremum of the potential, 
provides us with another real curve in the  ($\Delta_0, \varphi=\bar{\varphi}$) plane. 
The values ($\Delta_{0*}, \varphi_*=\bar{\varphi}_*$) are the intersection of these two. 
The schematic figure of the intersection is displayed in Figure \ref{intersection}. 
By tuning our original input functions, it is possible to arrange such intersection. 
Conversely, as an inverse problem, for given $\Delta_{0*}$ and the height of the $\Delta_0$ profile, 
one can always find the values of the coefficients in eq. (\ref{gap2}) and the coefficient function in eq. (\ref{7}) 
that accomplish this. 
Dynamical supersymmetry breaking has been realized. 

As for the former case, as in the latter case, we can safely replace eq. (\ref{phiflat1}) by 
\ba
\frac{V_0}{N^2 |m_s|^4} = \frac{\left( \frac{\partial \tilde{\delta}}{\partial \varphi} \right)}{2\partial (\ln |m_s|^2)} 
\left( \frac{\Delta_0 + \bar{\Delta}_0}{2} \right)^2 
+ \frac{\left( \frac{\partial \theta}{\partial \varphi} \right) |\Delta_0|^2}{2\partial(\ln|m_s|^2)}C.
\label{phiflat1final}
\ea
The values ($\Delta_{0*}, \bar{\Delta}_{0*}, \varphi_*, \bar{\varphi}_*$) can be determined by the intersection of eq. (\ref{gap1}) and eq. (\ref{phiflat1final}). 
We will not carry out the (numerical) analysis for this case further in this paper.

\subsection{determination of $F_*$}
Let us now turn to the analysis of the remaining equation of our variational problem, eq. (\ref{Fflat}). 
In our current treatment, 
\ba
F = -\frac{1}{g} \overline{W}' + \frac{1}{g} \frac{\partial}{\partial \bar{F}} V_{{\rm 1-loop}} \approx 
\left. -\frac{1}{g} \overline{W}' + \frac{1}{g} \frac{\partial}{\partial \bar{F}} V_{{\rm 1-loop}} \right|_{F=0}.
\ea
As the stationary values $(D_*, \varphi_*, \bar{\varphi}_*)$ are already determined, 
this equation and its complex conjugate determine $F_*$ and $\bar{F}_*$:
\ba
F_* = \frac{1}{g(\varphi_*, \bar{\varphi}_*)} \left( \left. -\overline{W}'(\varphi_*, \bar{\varphi}_*) 
+ \frac{\partial}{\partial \bar{F}} V_{{\rm 1-loop}}(D_*, \varphi_*, \bar{\varphi}_*, F, \bar{F}) \right|_{F=\bar{F}=0} \right).
\ea
Note that, knowing $V_{{\rm 1-loop}}$ explicitly in eq. (\ref{1loopeffpot}), 
the RHS can be evaluated. 
We can check the consistency of our treatment through $f_3$ in eq. (\ref{f3}) by $|f_3| \ll 1$.

\subsection{numerical study of the gap equation}

In this subsection, we study some numerical solutions to the gap equation eq. (\ref{gap2}) 
and the stationary condition for $\varphi$ eq. (\ref{7}) 
in the real $\Delta_0$ case. 
The equations we should study are
\ba
&&0 = 2 
\left( c' + \frac{1}{64 \pi^2} \right) 
+ \frac{1}{32\pi^2}
\left\{
\frac{\tilde{A}}{2} \Delta_{0*}^2 
\right. \nonumber \\
&& \left. \left. 
-\frac{1}{\sqrt{1 + \Delta_{0*}^2}} 
\left(
\lambda_0^{(+)3} \left( 2 \log \lambda_0^{(+)2} + 1 \right)
- \lambda_0^{(-)3} \left( 2 \log \lambda_0^{(-)2} + 1 \right)
\right) \right|_{\Delta_0=\Delta_{0*}}
\right\}, 
\label{gapatsp} \\
&&\frac{V_0}{N^2|m_{s*}|^4} = \frac{\left. \frac{\partial \delta(\varphi, \bar{\varphi})}{\partial \varphi} \right|_{\varphi_*, \bar{\varphi}_*}}{2 \partial (\ln |m_{s*}|^2)} \Delta_{0*}^2, 
\ea
where we note that $\delta(\varphi, \bar{\varphi})$ in the gap equation (\ref{gap2}) vanishes at the stationary point in the real $\Delta_0$ case.  
By using eqs. (\ref{realpotential}) and (\ref{delta}), the second condition can be rewritten after dividing by $\Delta_{0*}^2$ as
\ba
&&\left(c' + \frac{1}{64\pi^2} \right) 
+ \frac{1}{32\pi^2} 
\left[
\frac{\tilde{A}}{8} \Delta_{0*}^2 
- \left. \frac{1}{\Delta_{0*}^2}\left(  |\lambda_0^{(+)}|^4 \log |\lambda_0^{(+)}|^2 - |\lambda_0^{(-)}|^4 \log |\lambda_0^{(-)}|^2 \right) \right|_{\Delta_0=\Delta_{0*}}
\right] \nonumber \\
&=& \frac{1}{4N^2 \partial \ln |m_{s*}|^2} \frac{{\rm Im}({\cal F}_*'' + \Lambda)}{r_{0*}^2 | m_{s*} |^4} 
\left[
\partial \left. \ln {\rm Im}({\cal F}''+\Lambda) \right|_{\varphi_*, \bar{\varphi}_*} 
- \frac{\partial (\left. r_0 |m_s|^2)^2 \right|_*}{(r_{0*}^2 |m_{s*}|^4)^2}
\right]. 
\label{phiflatatsp}
\ea
The nontrivial solution $\Delta_{0*} \ne 0$ to the gap equation (\ref{gapatsp}) is found by some region of the
 parameters $c'$ and $\tilde{A}$, which was already done in \cite{imaru1}. 
This solution fixes the LHS of eq. (\ref{phiflatatsp}) and $\varphi_*$ is determined by solving eq. (\ref{phiflatatsp}) in principle. 
In order to find $\varphi_*$ explicitly, the form of the prepotential ${\cal F}$ and  that of the superpotential $W$ 
must be specified. 
Here, we take a simple prepotential and a superpotential of the following type (with some abuse of the notation): 
\ba
{\cal F} &=& \frac{c}{2N}  {\rm tr} \varphi^2 + \frac{1}{3!MN} {\rm tr} \varphi^3  \equiv \frac{1}{2} c \varphi^2 + \frac{1}{3!M} \varphi^3, 
\label{F} \\
W &=& \frac{m^2}{N} {\rm tr} \varphi + \frac{d}{3!N} {\rm tr} \varphi^3 \equiv m^2 \varphi + \frac{d}{3!} \varphi^3,
\label{W}
\ea
where $c, d$ are dimensionless constants while $m, M$ carry dimensions. 
In particular, $M$ is a cutoff scale of the theory.  
This prepotential is minimal for DDSB. 
As for the superpotential, 
at least two terms are required to be supersymmetric at tree level. 
We can take a quadratic term $\varphi^2$ instead of the cubic one, 
but in that case, RHS of eq. (\ref{phiflatatsp}) becomes singular because of $\partial \ln |m_s|^2=0$. 

Substituting these ${\cal F}$ and $W$ into eq. (\ref{phiflatatsp}), we obtain
\ba
&&\left(c' + \frac{1}{64\pi^2} \right) 
+ \frac{1}{32\pi^2} 
\left[
\frac{\tilde{A}}{8} \Delta_{0*}^2 
- \left. \frac{1}{\Delta_{0*}^2}\left(  |\lambda_0^{(+)}|^4 \log |\lambda_0^{(+)}|^2 - |\lambda_0^{(-)}|^4 \log |\lambda_0^{(-)}|^2 \right) \right|_{\Delta_0=\Delta_{0*}}
\right] \nonumber \\
&=& -\frac{{\rm Im}(c + \Lambda)({\rm Im}~c)^4}{N^2} \frac{1}{\left( d \varphi_*/M \right)^2},
\label{phiflatfinal}
\ea
where we utilized the fact that $1/M, d, \varphi_*$ are real and $c$ is pure imaginary, which are necessary for $\Delta_0=\bar{\Delta}_0$. 
If we take the coefficients $c=i, d=1$ for further simplification, 
we can easily obtain a solution by tuning $N$ and ${\rm Im}\Lambda$.  
We note $0 \le \varphi_*/M \le 1$ for our effective theory to be valid. 
In our analysis carried out in this paper, we consider the region where the magnitude of the F-term is smaller compared to that of the D-term. 
Therefore, we need to check whether our solutions satisfy this property consistently. 
Let us consider the ratio of the auxiliary fields:  
\ba
\left| \frac{F_*}{D_*} \right| &=& \left| \frac{-g^{-1} \overline{W}'(\bar{\varphi}_*) 
+ g^{-1} \left. \frac{\partial}{\partial \bar{F}} V_{{\rm 1-loop}}(D_*, \varphi_*, \bar{\varphi}_*, F, \bar{F}) \right|_{F=\bar{F}=0}}{\Delta_{0*}/r_{0*}} \right| \nonumber \\
&=& \left| 
\frac{1}{\sqrt{2}\Delta_{0*} \frac{\varphi_*}{M}} 
\left[
\left( \frac{m}{M} \right)^2 + \frac{1}{2} \left( \frac{\varphi_*}{M} \right)^2
\right] 
\right. \nonumber \\
&& \left. 
+ i \frac{N^2}{\sqrt{2}\Delta_{0*}} \left( \frac{\varphi_*}{M} \right)^2
\left[
\frac{\tilde{A}}{128\pi^2} \Delta_{0*}^2
-\frac{1}{32\pi^2} (|\lambda_{0*}^+|^4 \log |\lambda_{0*}^+|^2 + |\lambda_{0*}^-|^4 \log |\lambda_{0*}^-|^2 +1) 
\right. \right. \nonumber \\
&& \left. \left. 
+\frac{1+ \frac{\Delta_{0*}^2}{2}}{32\pi^2 \sqrt{1+\Delta_{0*}^2}} 
\left\{
(\lambda_{0*}^+)^3 \left(\log |\lambda_{0*}^+|^2 + \frac{1}{2} \right) - (\lambda_{0*}^-)^3 \left(\log |\lambda_{0*}^-|^2 + \frac{1}{2} \right) 
\right\} 
\right]
\right|,
\ea
where the form of the prepotential and that of the superpotential in eq. (\ref{F}) and eq. (\ref{W}) are assumed
 and we have put $c=i, d=1$ in the second equality. 

Now, the numerical solutions to the gap equation and the stationary condition for $\varphi$ are listed in Table \ref{numerical}. 
In these examples, we have taken some values of $-\frac{N^2}{{\rm Im}(i + \Lambda)}$ and $m$ just for an illustration and 
the ratio $|F_*/D_*|$ and $|f_{3*}|$ are evaluated. 
We can find that the $F$-term is smaller than the $D$-term in some of these examples. 
\begin{table}[htb]
\begin{center}
  \begin{tabular}{|c|c|c|c|c|c|} 
  \hline
$c'+\frac{1}{64\pi^2}$ & $\tilde{A}/(4\cdot32\pi^2)$ & $\Delta_{0*}$  
& $\varphi_*/M~(-\frac{N^2}{{\rm Im}(i+\Lambda)})$ & $|F_*/D_*|$ & $|f_{3*}|$ \\ 
\hline
0.002 & 0.0001 & 0.477 & 0.707~(10000) & 2.621~($m=M$) & 1.77 \\
0.002 & 0.0001 & 0.477 & 0.707~(10000) & 0.524~($m \ll M$) & 0.35 \\
0.002 & 0.0001 & 0.477 & 0.707~(10000) & 0.860~($m=0.4M$) & 0.58 \\
\hline
0.003 & 0.001 & 1.3623 & 0.8639~(2000) & 0.825~($m=M$) & $>$1 \\
0.003 & 0.001 & 1.3623 & 0.8639~(2000) & 0.224~($m \ll M$) & 0.43 \\
0.003 & 0.001 & 1.3623 & 0.5464~(5000) & 1.092~($m=M$) & $>$1 \\
0.003 & 0.001 & 1.3623 & 0.5464~(5000) & 0.142~($m \ll M$) & 0.27 \\
0.003 & 0.001 & 1.3623 & 0.5464~(5000) & 0.911~($m=0.9M$) & 1.76 \\
0.003 & 0.001 & 1.3623 & 0.3863~(10000) & 1.444~($m=M$) & $>$1 \\
0.003 & 0.001 & 1.3623 & 0.3863~(10000) & 0.100~($m \ll M$) & 0.19 \\
0.003 & 0.001 & 1.3623 & 0.3863~(10000) & 0.960~($m=0.8M$) & 1.85 \\
 \hline
  \end{tabular}
\end{center}
\caption{Samples of numerical solutions for the gap equation and the stationary condition for $\varphi$. 
The ratio $|F_*/D_*|$ and $|f_{3*}|$ are also evaluated for consistency check.}
\label{numerical}
\end{table}

\subsection{second variation of the potential and the mass of the scalar gluons}

We now turn to the question of the second variations of the scalar potential 
\ba
V_{{\rm scalar}} = V(D=D_*(\varphi, \bar{\varphi}), F=F_*(\varphi, \bar{\varphi}) \approx 0, \bar{F}=\bar{F}_*(\varphi, \bar{\varphi}) \approx 0, \varphi, \bar{\varphi}) 
\ea
at the stationary point ($D_*(\varphi_*,\bar{\varphi}_*), 0, 0, \varphi_*, \bar{\varphi}_*$). 
It is convenient to separate $V(D, F, \bar{F}, \varphi, \bar{\varphi})$ into two parts: 
\ba
V = {\cal V}+V_0. 
\ea
Here
\ba
{\cal V}(F, \bar{F}, \varphi, \bar{\varphi}) &\approx& - g F \bar{F} - F W' - \bar{F} \bar{W}' + (\partial_F V_{{\rm 1-loop}})_* F + (\partial_{\bar{F}} V_{{\rm 1-loop}})_* \bar{F} \nonumber \\
&&+ \frac{1}{2} (\partial_F^2 V_{{\rm 1-loop}})_* F^2 + \frac{1}{2} (\partial^2_{\bar{F}} V_{{\rm 1-loop}})_* \bar{F}^2 + (\partial_F \partial_{\bar{F}} V_{{\rm 1-loop}})_* F \bar{F},
\label{calV}
\ea
and 
\ba
V_0(D, \varphi, \bar{\varphi}) = V(D, \varphi, \bar{\varphi}, F=0, \bar{F}=0). 
\label{V0}
\ea
In eq. (\ref{calV}), we have extracted   the $F, \bar{F}$ dependence of $V_{{\rm 1-loop}}$ (eq. (\ref{1loopeffpot})) 
as series and $_*$ indicates that they are evaluated at $(D_*, \varphi_*, \bar{\varphi}_*, 0, 0)$ after the derivatives are taken. 
Eq. (\ref{V0}) has been computed in eq. (\ref{renpotential}) and eq. (\ref{realpotential}). 
We will compute the second partial derivatives and the second variations of $V_{{\rm scalar}}$, 
using the formula in the appendix. 

For ${\cal V}$, $\vec{y}_L = (F, \bar{F}), \vec{y}_R=(\varphi, \bar{\varphi})$, 
\ba
&&M_{RR_*} \equiv 
\left(
\begin{array}{cc}
\partial^2 {\cal V}, & \partial \bar{\partial} {\cal V} \\
\bar{\partial} \partial {\cal V}, & \bar{\partial}^2 {\cal V} \\
\end{array}
\right)_* \approx 0, \\
&&M_{RL_*} \equiv 
\left(
\begin{array}{cc}
\partial \partial_F {\cal V}, & \partial \partial_{\bar{F}} {\cal V} \\
\bar{\partial} \partial_F {\cal V}, & \bar{\partial} \partial_{\bar{F}} {\cal V} \\
\end{array}
\right)_* \approx 
\left(
\begin{array}{cc}
-W''+ (\partial \partial_F V_{{\rm 1-loop}}), & (\partial \partial_{\bar{F}} V_{{\rm 1-loop}}) \\
(\bar{\partial} \partial_F V_{{\rm 1-loop}}), & -\overline{W}'' + (\bar{\partial}\partial_{\bar{F}} V_{{\rm 1-loop}}) \\
\end{array}
\right)_*, \\
&&M_{LR_*} = M_{RL_*}^t, \\
&&M_{LL_*} \equiv 
\left(
\begin{array}{cc}
\partial_F^2 {\cal V}, & \partial_F \partial_{\bar{F}} {\cal V} \\
\partial_{\bar{F}} \partial_F {\cal V}, & \partial_{\bar{F}}^2 {\cal V} \\
\end{array}
\right)_* \approx 
\left(
\begin{array}{cc}
(\partial_F^2 V_{{\rm 1-loop}}), & -g + (\partial_F \partial_{\bar{F}} V_{{\rm 1-loop}}) \\
-g + (\partial_{\bar{F}} \partial_F V_{{\rm 1-loop}}), & (\partial_{\bar{F}}^2 V_{{\rm 1-loop}}) \\
\end{array}
\right)_*.
\ea
Here we have denoted by $*$ that the derivatives are evaluated at the stationary point. 

We obtain, after some computation, 
\ba
\delta^2 {\cal V}_* &\approx& \frac{1}{2} \delta \vec{y}_R^t M_{RL_*} (-M_{LL_*}^{-1}) M_{LR_*} \delta \vec{y}_R 
\equiv \frac{1}{2} \delta \vec{y}_R^{\dagger}
\left(
\begin{array}{cc}
{\cal M}_{\varphi\bar{\varphi}} & {\cal M}_{\varphi\varphi} \\
{\cal M}_{\bar{\varphi}\bar{\varphi}} & {\cal M}_{\bar{\varphi}\varphi} \\
\end{array}
\right)_* \delta \vec{y}_R. 
\label{massmatrix}
\ea
Here
\ba
&&{\cal M}_{\varphi\bar{\varphi}} = \frac{1}{g(G^2 - |C|^2)} \left( G (|A|^2 + |B|^2) + C A \bar{B} + \bar{C} \bar{A} B \right), \\
&&{\cal M}_{\varphi \varphi} = \frac{1}{g(G^2 - |C|^2)} \left( 2G A B + C A^2 + \bar{C} B^2 \right), \\
&&G \equiv 1-\frac{\partial_F \partial_{\bar{F}}V_{{\rm 1-loop}}}{g}, \quad C \equiv \frac{\partial_{\bar{F}}^2 V_{{\rm 1-loop}}}{g}, \quad
A \equiv W'' -\partial \partial_F V_{{\rm 1-loop}}, \quad
B \equiv -\partial \partial_{\bar{F}} V_{{\rm 1-loop}}. \nonumber \\
\ea
Here in the last line of eq.(\ref{massmatrix}), we have changed the real quadratic form into the complex one.  
We see that in the region $|(\partial_F \partial_{\bar{F}} V)_0|_*, |(\partial_F^2 V)_0|_*, \ll g_*$, 
the matrix ${\cal M}_*$ is well approximated by
\ba
{\cal M}_* \approx \frac{1}{g} 
\left(
\begin{array}{cc}
|A|^2 + |B|^2, & 2AB \\
2 \bar{A}\bar{B}, & |A|^2 + |B|^2 \\
\end{array}
\right)_*. 
\ea
The two eigenvalues are 
\ba
\frac{1}{g} (|A| \pm |B|)_*^2 = \frac{1}{g} \left( |W''- (\partial \partial_F V_{{\rm 1-loop}})| \pm 
|(\partial \partial_{\bar{F}}V_{{\rm 1-loop}})| \right)_*^2, 
\ea
respectively, ensuring the positivity of (\ref{massmatrix}). 

For $V_0$, $y_L =D, \vec{y}_R=(\varphi, \bar{\varphi})$, 
\ba
&&M_{RR_*} = 
\left(
\begin{array}{cc}
\partial^2 V_0, & \partial \bar{\partial} V_0 \\
\bar{\partial} \partial V_0, & \bar{\partial}^2 V_0 \\
\end{array}
\right)_*, \qquad 
M_{RL_*} = 
\left(
\begin{array}{c}
\partial \partial_D V_0 \\
\bar{\partial} \partial_D V_0 \\
\end{array}
\right)_*, \\
&&M_{LR_*} =M_{RL}^*, \qquad M_{LL_*} = \partial_D^2 V_{0*}.  
\ea
We know that the $D$ profile of $V_0(D, \varphi, \bar{\varphi})$ near the stationary point is 
convex to the top and 
we fit this by
\ba
V_0 = V_h(\varphi, \bar{\varphi}) - \frac{\alpha(\varphi, \bar{\varphi})}{2} \left( D - D_*(\varphi, \bar{\varphi}) \right)^2 
+ {\cal O}((D - D_*(\varphi, \bar{\varphi}))^4). 
\ea
Here $\alpha$ is a positive real function of $\varphi, \bar{\varphi}$ and $V_h(\varphi, \bar{\varphi}) = V_0(D_*(\varphi, \bar{\varphi}), \varphi, \bar{\varphi})$. 
One can check
\ba
-M_{RL_*} M_{LL_*}^{-1} M_{LR*} = \alpha_* 
\left(
\begin{array}{c}
\partial D_*\\
\bar{\partial} D_*\\
\end{array}
\right)_*
((\partial D_*), (\bar{\partial} D_*))_*,
\ea
while
\ba
M_{RR*} = 
\left(
\begin{array}{cc}
\partial^2 V_h & \partial \bar{\partial} V_h \\
\bar{\partial} \partial V_h & \bar{\partial}^2 V_h \\
\end{array}
\right)_*
-\alpha 
\left(
\begin{array}{cc}
(\partial D_*)^2 & |\partial D_*|^2 \\
|\partial D_*|^2 & (\bar{\partial} D_*)^2 \\
\end{array}
\right)_*
\ea
and
\ba
\delta^2 V_{0*} &=& \frac{1}{2} \delta \vec{y}_R 
\left( M_{RR*} - M_{RL*} M_{LL*}^{-1} M_{LR*} \right)_*
\delta \vec{y}_R \nonumber \\
&=& \delta \vec{y}_R^\dag 
\left(
\begin{array}{cc}
\partial \bar{\partial} V_h & \partial^2 V_h \\
\bar{\partial}^2 V_h & \partial \bar{\partial} V_h \\
\end{array}
\right)_*
\delta \vec{y}_R 
\equiv \delta \vec{y}_R^\dag {\cal M}_{h*} \vec{y}_R. 
\label{V0massmatrix}
\ea

\begin{figure}[htbp]
 \begin{center}
  \includegraphics[width=80mm]{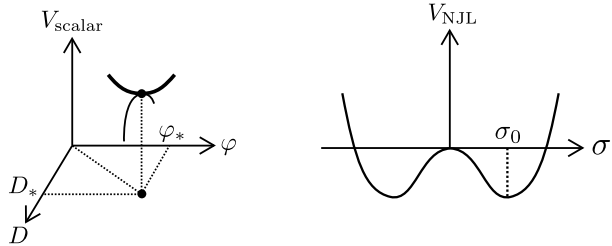}
 \end{center}
 \caption{Comparison of $V_{{\rm scalar}}$ around the stationary value $(D_*, \varphi_*)$ with $V_{{\rm NJL}}$.} 
 \label{potential}
\end{figure}

  The entire contribution of the second variation  
  $\delta^2 V_* = \delta^2 {\cal V}_* + \delta^2 V_{0*}$ to the leading order in the Hartree-Fock approximation 
  is given by eqs.(\ref{massmatrix}), (\ref{V0massmatrix}). 
  The mass of the scalar gluons squared is obtained by multiplying the combined mass matrix by $g^{-1}_*$:
\ba
  g^{-1}_* ({\cal M}_* + {\cal M}_{h*}), 
\ea
generalizing the tree formula.  
  In practice, we just need a well-approximated formula  valid in the region we work with and one can invoke
  the $U(1)$ invariance to ensure that the two eigenvalues of the complex scalar gluons are degenerate. 
  Let us, therefore, use the expression
\ba
&&\frac{1}{g} |W''- (\partial \partial_F V_{{\rm 1-loop}})|_*^2  + \partial \bar{\partial} V_{h*} \nonumber \\
&=& \left|
\frac{\varphi_*}{M} 
-iN^2 \left( \frac{\varphi_*}{M} \right)^2 
\left[
\frac{A(\varepsilon, \gamma)}{32\pi^2} 
\left\{
(\lambda_{0*}^{+})^4 + (\lambda_{0*}^{-})^4 + \frac{2}{\varphi_*/M} -2 -\frac{3}{4} \Delta_{0*}^2 +\frac{1}{8} \Delta_{0*}^4
\right\} \right. \right. \nonumber \\
&& \left. \left. -\frac{1}{32\pi^2} 
\left\{
(\lambda_{0*}^{+})^4 \log (\lambda_{0*}^{+})^2 + (\lambda_{0*}^{-})^4 \log (\lambda_{0*}^{-})^2 
\right. \right. \right. \nonumber \\
&& \left. \left. \left. + \frac{3(1+\frac{\Delta_{0*}^2}{2})}{\sqrt{1+\Delta_{0*}^2}} 
\left((\lambda_{0*}^{-})^3 \left( \log (\lambda_{0*}^{-})^2 + \frac{1}{2} \right) 
- (\lambda_{0*}^{+})^3 \left( \log (\lambda_{0*}^{+})^2 + \frac{1}{2} \right) \right)
\right\} 
-\frac{2}{32\pi^2} \frac{1}{\varphi_*/M}
\right] \right|^2 M^2 \nonumber \\
&& + 2N^2 \left(\frac{\varphi_*}{M} \right)^2 
\left[
2\left( c' +\frac{1}{64\pi^2} \right) \Delta_{0*}^2 
+ \frac{2}{32\pi^2} \left( \frac{\tilde{A}}{8} -(\lambda_{0*}^+)^4 \log (\lambda_{0*}^+) - (\lambda_{0*}^-)^4 \log(\lambda_{0*}^-)^2 \right) 
\right. \nonumber \\
&& \left. 
+ \frac{{\rm Im}(i+\Lambda)}{N^2} \frac{1}{\varphi_*/M}
\right] M^2
\label{stabcriterion}
\ea
to check the local stability of the potential and the mass. 
The above expression is obtained for our simple example of ${\cal F}$ and $W$
\ba
{\cal F} = \frac{i}{2} \varphi^2 + \frac{1}{3!M} \varphi^3, \qquad
W = m^2 \varphi + \frac{1}{3!} \varphi^3,
\label{W}
\ea 
and the real case $\Delta_0=\bar{\Delta}_0$ is applied. 
Using the numerical analyses carried out in the last subsection, 
we have made a list of data on  eq. (\ref{stabcriterion}). 
\begin{table}[htb]
\begin{center}
  \begin{tabular}{|c|c|c|c|l|} 
  \hline
$c'+\frac{1}{64\pi^2}$ & $\tilde{A}/(4\cdot32\pi^2)$ & $\Delta_{0*}$  
& $\varphi_*/M~(-\frac{N^2}{{\rm Im}(i+\Lambda)})$ & scalar gluon mass \\ 
\hline
0.002 & 0.0001 & 0.477 & 0.707~(10000) & 0.4998 + 0.0056 $N^2$ + $8.607 \times 10^{-7} N^4$ \\
\hline
0.003 & 0.001 & 1.3623 & 0.8639~(2000) & 0.7463 + 0.0106 $N^2$ + $2.653 \times 10^{-4} N^4$  \\
0.003 & 0.001 & 1.3623 & 0.5464~(5000) & 0.2986 + 0.0008 $N^2$ + $4.694 \times 10^{-5} N^4$ \\
0.003 & 0.001 & 1.3623 & 0.3863~(10000) & 0.1492 $-$ 0.0024 $N^2$ +$7.235 \times 10^{-5} N^4$ \\
 \hline
  \end{tabular}
\end{center}
\caption{Samples of numerical values for the scalar gluon masses.}
\label{sgluonmass}
\end{table}
Except for the last case in the table \ref{sgluonmass}, 
the scalar gluon masses squared are found to be positive for any $N$, 
which implies that our stationary points are locally stable. 
Even in the last case, the stability is ensured for small $N$. 
In these data, we have checked that 
the inequalities $|(\partial_F \partial_{\bar{F}} V_{{\rm 1-loop}})|_*, |(\partial_F^2 V_{{\rm 1-loop}})|_* \ll g_*$ are in fact  satisfied. 
As a summary of our understanding, 
a schematic figure is drawn in Fig. \ref{potential}, which illustrates the local stability of the scalar potential 
at the vacuum of dynamically broken supersymmetry in comparison with the well-known NJL potential.

\subsection{summary and choice of regularization and subtraction scheme} 
 In this paper, we have considered the theory specified by the general  ${\cal N}=1$ supersymmetric lagrangian  eq. (\ref{KtauW}), 
have regularized the theory by the supersymmetric dimensional regularization (dimensional reduction) 
and have subtracted the part of the $1/(\epsilon)$ poles of the regularized one-loop effective action in eq. (\ref{1loopeffpot}) 
by the supersymmetric subtraction scheme defined by the condition  eq. (\ref{rencond}).
The upshot is an effective potential eqs. (\ref{renpotential2}), (\ref{realpotential}) as a function of the background constant scalar 
and the order parameter $D$ of supersymmetry, with another order parameter $F$ of supersymmetry being induced and treated perturbatively. 
Supersymmetry is dynamically broken as is represented by the non-vanishing value of the order parameters at the stationary point. 
The original infinity is transmuted into the infinite constant $\Lambda$ 
which is the coefficient of the counterterm and the effective potential has been recast to describe 
the behavior of the theory well below the UV cutoff residing in the prepotential function. 
As the theory is perturbatively nonrenormalizable, 
$\Lambda$ is still present in our final expressions of the effective potential and we regard it to take a large value.

We now make brief comments on other regularizations and subtraction schemes which we did not employ in this paper. 
The relativistic momentum cutoff is a natural choice of the NJL theory as we mentioned earlier 
but regularizing the integral eq. (\ref{1loopeffaction}) by the momentum cuoff leads us to a rather unwieldy expression. 
See  ref. \cite{imaru2}. 
Unlike supersymmetric dimensional reduction \cite{Siegel},  
the momentum cutoff ${\it per se}$, while preserving the equality between the bose and fermi degrees of freedom, 
does not have a firm basis on the regularized action which the supersymmetry algebra acts on. 
Moreover, as is clear from $(A.1)$ of ref. \cite{imaru2}, the result violates the positivity of the effective potential 
in the vicinity of the origin in the $\Delta$ profile. 
This violation is a necessity in the broken chiral symmetry of the NJL theory 
but here it contradicts with the positive semi-definiteness of energy that the rigid supersymmetric theory possesses. 
Turning to the choice of the subtraction scheme, 
one might also like to apply the ``(modified) minimal subtraction scheme" in our one-loop integral eq. (\ref{1loopeffpot}). 
While we do not know how to justify this prescription here, the subsequent analyses proceed almost in the same way 
and the main features of the equations obtained from our variational analyses and the conclusions are unchanged.

\section*{Acknowledgements}
We would like to thank Kenzo Inoue, Tatsuo Kobayashi, Ken-ichi Okumura for valuable discussions. 
The authors' research is supported in part by the Grant-in-Aid for Scientific Research  
  from the Ministry of Education, Science and Culture, Japan (23540316 (H.~I.), 24540283 (N.~M.))
  and by Keio Gijuku Academic Development Funds (N.~M.).

\newpage

\appendix
\section{NJL effective action}

In this appendix, we briefly recall a few aspects of the $SU(N)$ Nambu-Jona-Lasinio model. 
\ba
{\cal L}_{{\rm NJL}} = \bar{\Psi} i \partial \!\!\! / \Psi 
+ \frac{\lambda/N}{2} 
\left[
(\bar{\Psi} \Psi)^2 + (\bar{\Psi} i \gamma_5 \Psi)^2
\right]
\label{NJLlag}
\ea
The equivalent lagrangian is
\ba
{\cal L} &=& {\cal L}_{{\rm NJL}} -\frac{1}{2} \frac{1}{\lambda/N} 
\left[
\left( \sigma + \frac{\lambda}{N} \bar{\Psi} \Psi \right)^2 
+ \left( \pi + \frac{\lambda}{N} \bar{\Psi} i \gamma_5 \Psi \right)^2
\right] \nonumber \\
&=& -\frac{1}{2} \frac{1}{\lambda/N} \sigma^2 -\frac{1}{2} \frac{1}{\lambda/N} \pi^2 
+ \bar{\Psi} \left( i \partial \!\!\!/ - \sigma -i\gamma_5 \pi \right) \Psi. 
\ea
The 1PI vertex function (or the effective action) $\Gamma_{{\rm 1PI}}[\sigma, \pi]$ to one-loop (or leadig order in $1/N$) reads
\ba
i \Gamma_{{\rm 1PI}}[\sigma, \pi] = -\frac{i}{2} \frac{1}{\lambda/N} \int d^4x \left( \sigma^2 + \pi^2 \right) 
+ N \ln {\rm Det} \left( i \partial \!\!\!/ -\sigma(x) -i\gamma_5 \pi(x) \right). 
\ea
The gap equation is 
\ba
0 &=& \left. \frac{i \delta \Gamma_{{\rm 1PI}}}{\delta \sigma(x)} \right|_{\sigma(x) = \langle \sigma \rangle =\sigma_0, \pi(x)=0} 
= -i \frac{1}{\lambda/N} \sigma_0 - N \int \frac{d^4k}{(2\pi)^4} {\rm Tr} \frac{1}{k\!\!\!/ - \sigma_0} \nonumber \\
&=& \sigma_0 \left(-i \frac{1}{\lambda/N}  - 4N \int \frac{d^4k}{(2\pi)^4} \times \frac{(-1)}{k^2 - \sigma_0^2} \right). 
\label{gapeq}
\ea

\section{formula for the second variation}
In this appendix, we recall the formula for the second variation of a multivariable function subject to a set of stationary constraints. 
Let $V$ be the function of two sets of variables: $\{\{y^1, \cdots, u^{n(L)} \}\}={\cal D}_L, \{\{ y^{n(L)+1}, \cdots, y^{n(L)+n(R)} \}\}={\cal D}_R$. 
Namely, 
\ba
V=V(y^1,\cdots, y^{n(L)}, y^{n(L)+1}, \cdots, y^{n(L)+n(R)})
\ea
under
\ba
\frac{\partial V}{\partial y^i} =0, i=1, \cdots, n(L). 
\label{stationary}
\ea
Let the second variation of V be
\ba
\delta^2 V \equiv \frac{1}{2} \sum_{y^i, y^j \in {\cal D}_L \cup {\cal D}_R} \frac{\partial^2 V}{\partial y^i \partial y^j} \delta y^i \delta y^j
\label{2ndvariation}
\ea
but $\delta y^i \in {\cal D}_L$ are not independent variations.

It is convenient to introduce a new vector notation:
\ba
\vec{y}_L = (y^1, \cdots, y^{n(L)})^t, \quad \vec{y}_R = (y^{n(L)+1}, \cdots, y^{n(L)+n(R)})^t, e.t.c
\ea
Define
\ba
M_{X, X'} = 
\left(
\frac{\partial^2 V}{\partial y^i \partial y^j}
\right),\quad
y^i \in {\cal D}_X, \quad y^j \in {\cal D}_{X'}~X, X' ~{\rm are~either}~L~{\rm or}~R. 
\ea
Eq.(\ref{2ndvariation}) reads
\ba
\delta^2 V = \frac{1}{2} (\delta \vec{y}_R, M_{RR} \delta\vec{y}_R) 
+ (\delta \vec{y}_R, M_{RL} \delta\vec{y}_L)
+ \frac{1}{2} (\delta \vec{y}_L, M_{LL} \delta\vec{y}_L)
\label{second}
\ea
Varying (\ref{stationary}) with respect to $\vec{y}_L$ and $\vec{y}_R$, 
we obtain
\ba
M_{LL} \delta \vec{y}_L + M_{LR} \delta \vec{y}_R = 0. 
\ea
Hence
\ba
\delta \vec{y}_L = - M_{LL}^{-1} M_{LR} \delta \vec{y}_R. 
\ea
Substituting this into (\ref{second}), we obtain
\ba
\delta^2 V = \frac{1}{2} \left( \delta\vec{y}_R, \left( M_{RR} - M_{RL}M_{LL}^{-1}M_{LR} \right) \delta \vec{y}_R \right). 
\ea
The generic scalar mass matrix in the text can be read off $M_{RR} - M_{RL}M_{LL}^{-1}M_{LR}$ at the stationary value.

\end{document}